\documentclass[%
preprint,%
amssymb, amsmath,%
pre]{revtex4-2}
\usepackage{natbib}%
\usepackage{url}
\usepackage{textcase}
\usepackage{bm}%
\usepackage{here}%
\usepackage{color}
\usepackage{atbegshi}
\usepackage[unicode=true,bookmarks=true]{hyperref}
\usepackage{bookmark}
\usepackage{graphicx}

\hypersetup{
 pdfauthor={makoto},
 pdftitle={},
 pdfkeywords={},
 pdfsubject={},
 pdfcreator={Emacs 26.3 (Org mode 9.3.7)}, 
 pdflang={English}}

\begin{document}
\preprint{aaa/bbb}
\title{Optimal external forces of the lock-in phenomena for the flow past inclined plate in a uniform flow}
\author{Makoto Iima}
\email{iima@hiroshima-u.ac.jp}
\affiliation{Graduate School of Integrated Life Sciences, Hiroshima University, 1-7-1, Kagamiyama Higashihiroshima, Hiroshima, 739-8521, Japan}
\date{\today}

\begin{abstract}
We theoretically studied the optimal control, frequency lock-in, and phase lock-in phenomena due to the spatially localized periodic forcing in the ﬂow past the inclined plate. Although frequency lock-in is evident in many ﬂuid phenomena, especially ﬂuid-structure interactions, not many researchers have investigated it using a theoretical approach based on ﬂow details. We obtained detailed information on the lock-in phenomena to external periodic forcing using phase reduction theory, a mathematical method for extracting the dynamics near the limit cycle. Furthermore, the optimal forces applied to the velocity ﬁeld were determined under the condition of the minimum forcing energy and maximum lock-in range. The study of uniform periodic forces applied within spatially conﬁned regions led us to conclude that the effective lock-in position, which includes both the upstream and downstream areas of the plate, depends on the principal frequency of the force. The frequency lock-in range of these forces was analyzed and compared with theoretical predictions.
\end{abstract}

\pacs{aaa}

\maketitle

\section{Introduction}
Effective flow control is demanded in many research areas, e.g. fluid engineering, nonlinear physics, and environmental research.
In particular, the frequency lock-in and the phase lock-in under external periodic forcing
 have been investigated in the context of fluid-structure interactions.
The examples of such interactions are spring-suspended airfoils in transonic flows in terms of aircraft vibration due to the shock wave oscillation
\cite{gao17_mechan_frequen_lock_in_trans_buffet_flow,%
raveh11_frequen_lock_in_phenom_oscil%
},
 spring-suspended cylinders exerted by random waves 
 in terms of fatigue and failures of structures in offshore systems
\cite{abroug22_frequen_phase_lock_in_behin}
,
 and various problems in flow-induced vibrations
\cite{williamson04_v_ortex_i_nduced_v_ibrat,%
langre06_frequen_lock_in_is_caused%
}
.

When an external periodic force is applied, the lock-in details can be provided through laboratory experiments and time evolution of computational fluid dynamics.
It has been, however, difficult to determine the optimal form of the external force to achieve the lock-in phenomena
 even if the forces are weak,
 as it requires complete information of the flow response to external perturbations.
Thus, it would be very helpful if we could design the external force to be considered as a control input.

A mathematical tool called the phase reduction theory can be used for this purpose.
It can be applied to a dynamical system with a limit cycle (LC) and
 describes the essential dynamics near the LC.
The reduced equation (called phase equation) has few degrees of freedom
\cite{Kuramoto1984}
.
Phase reduction theory has been successfully applied to various rhythmic phenomena
\cite{pikovsky2001synchronization}
 in
 mechanical vibration (synchronization of metronomes
 \cite{pantaleone02_synch_metron}),
 ethology (synchronization of flashing fireflies
\cite{%
buck35_synch_flash_firef_exper_induc,%
ermentrout91_adapt_model_synch_firef_pterop_malac%
}),
 and biology (circadian rhythms
\cite{%
bell-pedersen05_circad_rhyth_from_multip_oscil,%
asgari_targhi18_mathem_model_circad_rhyth%
}), etc.
Compared to the applications in mechanical engineering and life sciences,
 its applications to fluid mechanics are under development;
 e.g.
 thermal convection
\cite{%
kawamura14_noise_induc_synch_oscil_convec_its_optim,%
kawamura19_phase_reduc_limit_torus_solut%
},
 K\'arm\'an's vortex street
\cite{%
taira18_phase_respon_analy_synch_period_flows,%
iima19_jacob_free_algor_to_calcul,%
khodkar20_phase_synch_proper_lamin_cylin,%
khodkar21_phase_lockin_lamin_wake_to,%
loe21_phase_reduc_synch_oscil_flow,%
nair21_phase_based_contr_period_flows%
}
, wake on a wing
\cite{%
iima21_phase_reduc_techn_target_region,%
nair21_phase_based_contr_period_flows%
},
 and other phenomena
\cite{%
skene21_phase_reduc_analy_period_therm,%
ricciardi22_trans_inter_phase_inter_effec%
}.

In the phase reduction theory,
 complete information of the phase response can be obtained from the phase sensitivity function
 (PSF)
\cite{%
Kuramoto1984,%
nakao15_phase_reduc_approac_to_synch_nonlin_oscil%
}.
Thus far, three techniques have been proposed to calculate the PSF, i.e.,
 the direct method
\cite{%
taira18_phase_respon_analy_synch_period_flows,%
khodkar20_phase_synch_proper_lamin_cylin,%
khodkar21_phase_lockin_lamin_wake_to,%
loe21_phase_reduc_synch_oscil_flow,%
nair21_phase_based_contr_period_flows%
},
the adjoint method
\cite{%
kawamura14_noise_induc_synch_oscil_convec_its_optim,%
kawamura19_phase_reduc_limit_torus_solut%
}, and the Jacobian-free projection method
\cite{%
iima19_jacob_free_algor_to_calcul,
iima21_phase_reduc_techn_target_region,%
novicenko11_comput_phase_respon_curves_via%
}.
The theoretical background of these methods can be found in Refs.
\cite{nakao15_phase_reduc_approac_to_synch_nonlin_oscil,%
iima21_phase_reduc_techn_target_region%
}.

The direct method measures the phase shift due to perturbation by time evolution and is suitable for cases where phase shifts due to a small number of degrees of freedom are of interest.
However, the accuracy of the phase-shift measurement is limited by the time step of the numerical calculation and requires a sufficiently long time for convergence.
The adjoint method obtains the PSF using time evolution of the adjoint equation derived from the equation.
This provides a convenient computational procedure, although the derivation of the adjoint equation is not always possible.
The Jacobian-free projection method can be used to obtain the PSF by computing the eigenvector of the matrix constructed by time evolution alone.
It can be applied to the system from which the adjoint equation is difficult to derive, although the limit cycle solution is required and the computational cost is higher than the adjoint method (yet lower than the direct method).

In the lock-in phenomena due to the periodic external forcing,
 the phase reduction theory can provide predictions, such as
 the frequency range of the external forcing for the frequency lock-in and the phase difference for the phase lock-in
\cite{nakao15_phase_reduc_approac_to_synch_nonlin_oscil}.
Furthermore, the optimal form of the external forcing under various conditions can be calculated
 as a constrained optimization problem
\cite{%
harada10_optim_wavef_entrain_weakl_forced_oscil,%
zlotnik12_optim_entrain_neural_oscil_ensem,%
zlotnik13_optim_wavef_fast_entrain_weakl,%
tanaka14_optim_entrain_with_smoot_pulse,%
tanaka14_synch_limit_weakl_forced_nonlin_oscil%
}
.
The optimal forms of the external forcing for the K\'am\'an's vortex street were studied.
Khodokar and Taira calculated the largest lock-in region for a sinusoidal form $1+\sin \Omega t$ applied at a single point.
They found that the best point to be near the separation point
\cite{khodkar20_phase_synch_proper_lamin_cylin}
.
Khodokar et al. studied the case where the cylinder is moving in uniform flow
\cite{khodkar21_phase_lockin_lamin_wake_to}.
Loe et al. studied the synchronization between the wake behind a 2D cylinder in a tube and the vibration of elastic walls in a sinusoidal form
\cite{loe21_phase_reduc_synch_oscil_flow}.
The lock-in region was maximized when the perturbation occurred near the downstream end of the cylinder.

Here, we considered the optimal forcing of the temporally periodic form applied uniformly to a spatially confined region.
Based on the spatial distributions of the PSF for both the cylinder and the plate,
\cite{iima19_phase_resopse_and_flow,%
iima21_phase_reduc_techn_target_region%
},
 the phase shift property due to perturbation has a complex spatio-temporal structure.
This fact implies that the practical control of the flow based on simple control inputs,
 e.g., spatially uniform input within a confined area and temporally simple (e.g. sinusoidal) function rather than an optimized waveform,
 may provide us different the lock-in properties from the optimized input by pointwise designable function.
Such a study will provide us insights into the appropriate region and frequency, especially for fluid engineering. 

In this study, we investigated the lock-in phenomena of the flow past the inclined plate in a wind tunnel in two-dimensional space.
First, we analyzed the qualitative characteristics of the PSF, which are useful for the designing of the control input.
Specifically, the details of the phase response to the external forces were described.
These results were used to tackle the lock-in problem for a uniform periodic external force within a confined region.
We aimed to answer the following questions:
\begin{enumerate}
  \item Where is the optimum region for the lock-in ?
  \item How does this position depend on the principal frequency and the direction of the force ?
\end{enumerate} 
We showed that the optimal position depends on the principal frequency and the direction of the force.
The optimal position may be away from the surface of the plate.
These results were compared with the theoretical prediction, which provides
 the maximum lock-in range under the constant energy of a external force
 and the minimum energy.

\section{Method}
\label{sec:Method}
\subsection{Fluid dynamics}
\label{sec:Fluid dynamics}
\begin{figure}[h]
  \centering
  \includegraphics[width=0.9\textwidth]{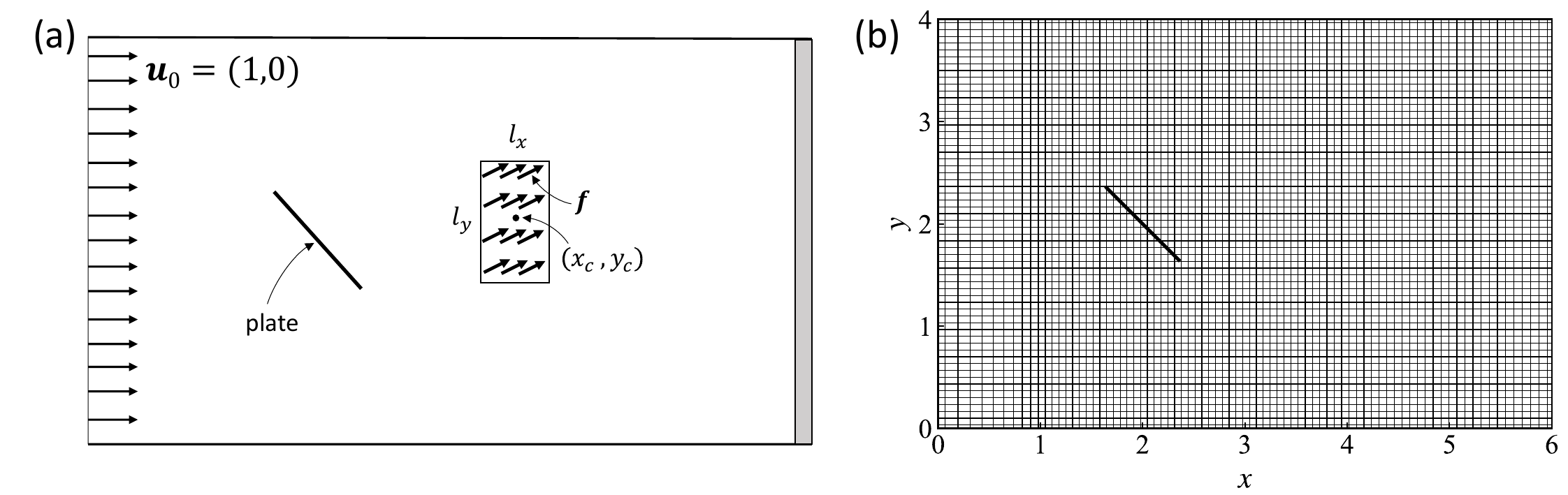}
  \caption{(a) Model Configuration. An inclined plate is placed in the wind tunnel. A uniform external force was applied within a rectangular area of size $l_x \times l_y$ centered at $(x_c, y_c)$. (b) Computational grid and the plate model. Grid lines are drawn on every other line for visibility. }
  \label{fig:configuration}
\end{figure}

The flow past a flat plate in a wing tunnel in two-dimensional space
 (Fig. \ref{fig:configuration}(a)) was considered. 
The flow is governed by the incompressible Navier-Stokes equations in a non-dimensional form:
\begin{eqnarray}
  \frac{\partial \bm{u}}{\partial t}
    + \bm{u}\cdot \nabla \bm{u}
    = -\nabla p + \frac{1}{Re} \Delta \bm{u}
    + \bm{f}(\bm{x}, \Omega t),
    \quad
    \nabla \cdot \bm{u} = \bm{0},
    \label{eq:Navier-Stokes eqs}
\end{eqnarray}
where $\bm{u}=(u,v)$ denotes the velocity, $p$ denotes the pressure, and $Re$ denotes the Reynolds number.
The time-periodic external force is denoted by $\bm{f}=(f_{x}, f_{y})$ and the angular frequency is denoted by $\Omega$.
The uniform flow is represented by $\bm{u}_0=(u_0, v_0)=(1,0)$. 
The system was assumed to be in a tunnel of width $4c$, where $c (=1)$ is the cord of the plate.
The system is non-dimensionalized by $c$ as the length scale and $c/u_0$ as the time scale;
 $Re=u_0 c/\nu$, where $\nu$ is the kinematic viscosity.

The computational domain was $[0,6c] \times [0,4c]$ to reduce the computational cost of calculating the functions describing the detailed phase response to an external force
 (Fig. \ref{fig:configuration}(b)).

The following boundary conditions were applied:
A constant velocity $\bm{u}_0$ was applied at the domain boundaries $x=0$, $y=0$, and $y=4c$.
The outflow boundary condition proposed by Dong et al.\cite{dong14_robus_accur_outfl_bound_condit},
 which aims at minimizing the domain truncation,
 was applied at the boundary $x=6c$.

In the following sections, the optimal external forces were considered.
In addition to the optimal forces predicted by the phase reduction theory
 (Sections. \ref{sec: Case A: Minimum energy that enables lock-in phenomena} and \ref{sec: Case B: Maximum lock-in region of frequency}),
  the optimal position was considered to maximize the frequency lock-in region
 under the condition of uniform external force within a rectangle of size $l_x \times l_y$ centered at $(x_c, y_c)$ (Fig. \ref{fig:configuration}(a)).

To solve Eqs. (\ref{eq:Navier-Stokes eqs}), a fractional step method was used.
The finite volume method was used for spatial discretization
\cite{liu98_numer_study_insec_fligh}. 
The Adams--Bashforth scheme and the Crank--Nicolson scheme were used for
 the time integration of the advection terms and that of the dissipation terms, respectively.
The flat plate was represented by an immersed boundary method\cite{uhlmann05_immer_bound_method_with_direc}.
The computational code was the same as that used in Ref. \cite{iima21_phase_reduc_techn_target_region}.

The center of the plate was set to $(x,y)=(2c,2c)$ and the angle of attack (AoA) was set to $\pi/4$.
The Reynolds number $Re$ was set to $200$.
An unequal and orthonormal grid was used, although the grid spacing in the region around the plate was uniform at $c/30$ (Fig. \ref{fig:configuration}(b)).
The number of grid points was $n_x \times n_y$ where $n_x=160$ and $n_y=120$.
In this setup, the periodic flow was achieved and the phase reduction theory can be applied.

The calculation scheme was compared with the spectral element method, in which 
 the computational domain was divided into quadrilateral elements, and physical quantities were represented using the spectral method\cite{blackburn19_semtex}; their results agreed reasonably well with each other (Appendix A).
In addition, we concluded that the phase sensitivity vector (see Section \ref{sec:Phase reduction theory}) near the plate was less sensitive to the size of the wind tunnel (Appendix B).

The periodic solution without external force was obtained numerically using the Newton--Raphson method
\cite{saiki07_numer_detec_unstab_period_orbit}
 under the condition where the relative errors of both the residue and the increment of the iteration
 were less than $10^{-10}$.
The period was $T=3.43958$ when a single period was segmented into $1408=2^7 \times 11$ time steps.
The origin of the phase was set as the time at which the maximum lift was attained.

The Jacobian-free projection method \cite{iima21_phase_reduc_techn_target_region} was used to obtain the projected phase-sensitivity function.
In this calculation, we focused on the response to the perturbation of the velocity components, $\bm{u}=(u,v)$ alone,
 whereas the responses to the pressure and variables in previous time steps were not calculated.
The Ritz value,
 an indicator for the convergence of the projection field
 \cite{iima21_phase_reduc_techn_target_region},
 was $6.59 \times 10^{-3}$ at the origin of the phase, which was reasonably small for the analysis.

\subsection{Phase reduction theory}
\label{sec:Phase reduction theory}
We analyzed an autonomous dynamical system with an external periodic force $\bm{F}(\Omega t)$:
\begin{equation}
  \frac{d\bm{X}}{dt} = \bm{G}(\bm{X}) + \bm{F}(\Omega t),
    \label{eq:dynamical systems}
\end{equation}
where $\bm{X} \in \mathbb{R}^{M}$ is the state in the $M-$dimensional phase space,
 $\bm{G}$ determines the autonomous dynamics system,
 and $\bm{F}(\Omega t)$ is the time-periodic external force with the angular frequency $\Omega$ and the period $T=2\pi/\Omega$,
 i.e., $\bm{F}(\Omega (t+T))=\bm{F}(\Omega t)$.
We assumed a weak external force to apply the phase reduction theory.
According to the phase reduction theory
\cite{Kuramoto1984,%
pikovsky2001synchronization%
}, the phase equation derived from Eq. (\ref{eq:dynamical systems}) reads:
\begin{equation}
  \frac{d\phi}{dt} = \omega + \bm{Z}(\phi) \cdot \bm{F}(\Omega t),
    \label{eq:phase equation}
\end{equation}
where $\phi \in [0,2\pi)$ is the phase, $\omega$ is the natural frequency, and
$\bm{Z}(\phi)$ is the phase sensitivity function.

We related $\bm{X}$ to the flow field data.
Suppose that the space is discretized by $n_x \times n_y$,
 the position $\bm{x}=(x,y)$ can be labeled by $m (=n_x n_y)$ indices
 $\bm{x}_1, \cdots, \bm{x}_{m}$.
In the same way, the velocity field $(u,v)$ is discretized to construct $\bm{X}$ as
\begin{equation}
  \bm{X}=(u_1,\cdots, u_{m}, v_1,\cdots, v_{m})
    \quad (M=2m),
    \label{eq:state vector}
\end{equation}
 where $u_j$ and $v_j$ are the values of $u$ and $v$ at $\bm{x}=\bm{x}_j$, respectively.
The external force applied to the fluid $\bm{f}(\bm{x}, \Omega t)$ is related to $\bm{F}(\Omega t)$ as 
\begin{equation}
  \bm{F}(\Omega t) = ( f_{x}(\bm{x}_1,\Omega t), \cdots, f_{x}(\bm{x}_m,\Omega t), f_{y}(\bm{x}_1,\Omega t), \cdots, f_{y}(\bm{x}_m,\Omega t)).
    \label{eq:relatioship betw. f_d and f}
\end{equation}

In the formal calculation, $\bm{X}$ contains more variables,
 $p$, and the variables that used in the numerical algorithm when the multistep method is used for time evolution
 (c.f. Section \ref{sec:Fluid dynamics}, Ref. \cite{iima21_phase_reduc_techn_target_region}).

The phase sensitivity vector, $\bm{q}(\bm{x}, \phi)=(q_u(\bm{x}, \phi),q_v(\bm{x}, \phi))$, describes the phase shift due to the unit force at the position $\bm{x}$ at the phase $\phi$.
The phase shift due to the perturbation $\Delta \bm{u}\, \delta(\bm{x}-\bm{x}_0)$,
 where $\Delta \bm{u}$
 and $\delta(\bm{x})$ represent
 a constant perturbation vector
 and the three-dimensional delta function, respectively,
 is expressed as $\Delta \bm{u} \cdot \bm{q}(\bm{x}_0)$
\cite{iima19_jacob_free_algor_to_calcul}
.

The relationship between $\bm{q}(\bm{x}, \phi)$ and $\bm{Z}(\phi)$ is:
\begin{eqnarray}
  \bm{Z}(\phi) = (q_u(\bm{x}_1, \phi)\Delta S_1, \cdots, q_u(\bm{x}_m,\phi)\Delta S_m, q_v(\bm{x}_1, \phi)\Delta S_1, \cdots, q_v(\bm{x}_m,\phi)\Delta S_m),
  \label{eq:relationship betw. Z and Q}
\end{eqnarray}
 where $\Delta S_j$ ($j=1,\cdots,m$) is the area allocated for the grid point $\bm{x}=\bm{x}_j$.

When the external force $\bm{F}(\Omega t)$ is weak, the phase equation is reduced to the following equation
 by the averaging over one period.
\begin{eqnarray}
  \frac{d\psi}{dt} &=& \Delta \omega + \Gamma(\psi),
    \label{eq:phase averaged equation}\\
  \Gamma(\psi) &=& \frac{1}{2\pi} \int_0^{2\pi} \bm{Z}(\theta + \psi) \cdot \bm{F}(\theta) d\theta
    (= \langle \bm{Z}(\theta + \psi) \cdot \bm{F}(\theta) \rangle),
    \label{eq:def of Gamma}
\end{eqnarray}
where $\psi = \phi - \Omega t$ is the phase difference between the system and the external force,
 and $\Delta \omega = \omega - \Omega$ is the frequency difference.
The function $\Gamma(\psi)$ is called the phase coupling function.

Scaling of variables scale in relation to the level of discretization, denoted as $m$, is examined below.
As both $\bm{q}_i(\bm{x}_k, \phi)$ and $f_j(\bm{x}_k, \Omega t)$ ($i=u,v;\;j=x,y;\;k=1,\cdots,m$) are independent of $m$,
 Eq.(\ref{eq:relationship betw. Z and Q}) and Eq.(\ref{eq:def of Gamma}) imply that
 $\bm{Z}(\phi) \simeq S/m \sim m^{-1}$ and $\Gamma(\psi) \sim m^0$, where $S=l_x l_y= \sum_{k=1}^{m} \Delta S_k$.
Therefore, the magnitude of $\bm{Z}(\phi)$ depends on the value of $m$ while $\Gamma(\phi)$ remains constant in the current formulation.

Equation (\ref{eq:phase averaged equation}) implies that the frequency lock-in occurs when
\begin{eqnarray}
  \Gamma_{\textrm{min}} < \Delta \omega < \Gamma_{\textrm{max}},\quad
  \Gamma_{\textrm{min}} = \textrm{min}_{0 \le \psi < 2\pi} \Gamma(\psi), \;\;
  \Gamma_{\textrm{max}} = \textrm{max}_{0 \le \psi < 2\pi} \Gamma(\psi).
    \label{eq:lock-in condition}
\end{eqnarray}

For later convenience, we defined
\begin{eqnarray}
  \psi_+ = \textrm{arg max} \Gamma(\psi), \quad
    \psi_- = \textrm{arg min} \Gamma(\psi).
    \label{eq:def of psi plus/minus}
\end{eqnarray}

For further analysis, we decomposed $\bm{f}(\bm{x}, \Omega t)$ and $\bm{q}(\bm{x}, \omega t)$ into Fourier series:
\begin{eqnarray}
  \bm{f}(\bm{x}, \Omega t) = \sum_{m=-\infty}^{\infty}\tilde{\bm{f}}(\bm{x};m) e^{im \Omega t}, \quad
    \bm{q}(\bm{x}, \omega t) = \sum_{m=-\infty}^{\infty}\tilde{\bm{q}}(\bm{x};m) e^{im \omega t},
    \label{eq:Fourier decompose of F and Q}
\end{eqnarray}  
where $\tilde{\bm{f}}$ and $\tilde{\bm{q}}$ are
 Fourier components of $\bm{f}$ and $\bm{q}$, respectively.
Similarly, we decomposed $\bm{F}(\phi)$ and $\bm{Z}(\phi)$ into Fourier series:
\begin{eqnarray}
  \bm{F}(\phi) = \sum_{m=-\infty}^{\infty}\tilde{\bm{F}}(m) e^{i m \phi}, \quad
  \bm{Z}(\phi) = \sum_{m=-\infty}^{\infty}\tilde{\bm{Z}}(m) e^{i m \phi}.
    \label{eq:Fourier decompose of fd and Z}
\end{eqnarray}  
Subsequently, Eq. (\ref{eq:def of Gamma}) provided the expression of $\Gamma(\psi)$ as
\begin{equation}
  \Gamma(\psi) = \sum_{m=-\infty}^{\infty}
    \tilde{\bm{Z}}(m) \cdot \tilde{\bm{F}}^*(m) e^{i m \psi},
    \label{eq:counpling function by Z(m) and f_d(m)}
\end{equation}
 where ${}^*$ represents the complex conjugate.

For $\bm{F}(t)= \epsilon \bm{F}_0 \sin(k t)$, $\Gamma(\psi) = -\epsilon |\bm{F}_0 \cdot \tilde{\bm{Z}}(k)| \sin(k \psi + \varphi)$,
 where
$\varphi = \textrm{arg} (\bm{F}_0 \cdot \tilde{\bm{Z}}(k))$.
Consequently, $\Gamma_{\textrm{max}} = \epsilon |\bm{F}_0 \cdot \tilde{\bm{Z}}(k)|$.
Furthermore, if certain components of $\bm{F}_0$ are zero,
 the corresponding components of $\tilde{\bm{Z}}(m)$ do not contribute to $\Gamma(\psi)$,
 which directly follows from Eqs.(\ref{eq:Fourier decompose of fd and Z}) and (\ref{eq:counpling function by Z(m) and f_d(m)}).

\subsection{Optimal external forces under several conditions}
\label{sec:Optimal external forces under several conditions}
\subsubsection{Case A: Minimum energy that enables lock-in phenomena}
\label{sec: Case A: Minimum energy that enables lock-in phenomena}
We considered an external force with minimum energy under the constraint of the lock-in phenomenon 
 based on Ref. \cite{zlotnik12_optim_entrain_neural_oscil_ensem}.
To obtain the optimal force, we minimize the Lagrangian function
$
  J_{\pm}[\bm{F}] = \langle |\bm{F}|^2 \rangle - \lambda (\Delta \omega + \Gamma(\psi_\pm)).
$
A straightforward calculation provides the minimizers $\bm{F}_{\pm}$ for $J_\pm$ as
\begin{equation}
  \bm{F}_{\pm}(\theta) = -\frac{\Delta \omega}{\langle Z^2 \rangle} \bm{Z}(\theta+\psi_{\pm}),
    \label{eq:optimal external force for minimum energy}
\end{equation}
 where the subscripts $+$ and $-$ correspond to the cases $\Omega > \omega$ and $\Omega < \omega$, respectively.
Thus, the external force with the minimum energy is proportional to the phase sensitivity function.

The energy of the external force, $P=\langle |\bm{F}_{\pm}|^2 \rangle$, is given by
$
  P= \frac{\Delta \omega^2}{\langle Z^2 \rangle}.
$
The coupling function $\Gamma_{\pm}(\psi)$, corresponding to $\psi_{\pm}$, respectively, is calculated as:
\begin{eqnarray}
  \Gamma_{\pm}(\psi) &=& \langle \bm{Z}(\theta + \psi) \cdot \bm{F}_{\pm}(\theta) \rangle
  = -\frac{\Delta \omega}{\langle Z^2 \rangle}
    \langle \bm{Z}(\theta + \psi) \cdot \bm{Z}(\theta + \psi_{\pm}) \rangle.
    \label{eq: Case A: coupling function}
\end{eqnarray}
The values of $\psi_{\pm}$ can be obtained by solving $\Gamma'(\psi_{\pm})=0$.
If we define the function as
\begin{eqnarray}
  g(x) &=&  \langle \bm{Z}'(\theta + x) \cdot \bm{Z}(\theta) \rangle,
    \label{eq:Definition of function g}
\end{eqnarray}
 the condition $\Gamma'(\psi_{\pm}) = 0$ is equivalent to $g(\psi -\psi_{\pm})=0$.
The following can be demonstrated:
\begin{equation}
  g(x+2\pi)=g(x), \quad g(-x) = -g(x),\quad g(0)=g(\pi)=0.
    \label{eq:property of g(x)}
\end{equation}
The first two identities are a consequence of the definition (\ref{eq:Definition of function g}), whereas the last identity is derived from the first two equations.
Any pair of $(\psi_{+}, \psi_{-})\;(\psi_+>\psi_-)$ that satisfied
\begin{equation}
  g(\Delta \psi)=0, \quad \Delta \psi = \psi_{+}-\psi_{-}
    \label{eq:Delta psi}
\end{equation}
constituted a valid solution.
In this study, $\psi_{-}=0$ was assumed.

\subsubsection{Case B: Maximum lock-in region of frequency}
\label{sec: Case B: Maximum lock-in region of frequency}
We considered the exeternal force that provides the maximum frequency lock-in region
 under the constraint of the constant energy, based on Ref. \cite{harada10_optim_wavef_entrain_weakl_forced_oscil}.
The lock-in range $R[\bm{F}]$ is defined as 
$
  R[\bm{F}]=\Gamma(\psi_+ )-\Gamma(\psi_-), \quad \langle |\bm{F}_{\pm}|^2 \rangle = P,
$
where $P$ is a constant.
The Lagrangian funtion is 
$
  J_{*}[\bm{F}]=R[\bm{F}] - \lambda (\langle |\bm{F}|^2 \rangle - P).
$
The minimizer $\bm{F}_{*}$ is:
\begin{equation}
  \bm{F}_{*} = \frac{1}{2\lambda}
    \left(
     \bm{Z}(\theta+\psi_+)-\bm{Z}(\theta+\psi_-)
    \right),
    \label{eq:minimizer for the external force of maximum lock-in range}
\end{equation}
 where the value of $\lambda$ is given by
$
  \lambda = \frac12 \sqrt{\frac{Q}{P}},\quad
    Q = \langle [\bm{Z}(\theta+\psi_+)-\bm{Z}(\theta+\psi_-)]^2 \rangle.
$
The coupling function $\Gamma(\psi)$ is:
\begin{eqnarray}
  \Gamma(\psi) &=&
    \frac{1}{2\lambda}
    \langle
    \bm{Z}(\theta+\psi) \cdot (\bm{Z}(\theta+\psi_+)-\bm{Z}(\theta+\psi_-))
    \rangle.\label{eq: Case B: coupling function}
\end{eqnarray}
The equation to determine $\psi_{\pm}$ is
\begin{equation}
g(\psi-\psi_+)- g(\psi-\psi_-) = 0,  
\end{equation}
 which is obtained by $\Gamma'(\psi)=0$.
Property (\ref{eq:property of g(x)}) gives that
\begin{equation}
  g(\Delta \psi) = 0.
\end{equation}

\subsubsection{Case C: Uniform force in the spatially localized area}
\label{sec: Case C: Uniform force in the spatially localized area}
In this paper, we have placed particular emphasis on the lock-in phenomena induced by a time-periodic external force,
 denoted as $\bm{f}(\bm{x}, \Omega t)$,
 which is spatially uniform and acts within a rectangular area of size $l_x \times l_y$
 centred in $\bm{x}_c=(x_c,y_c)$.
Specifically, this is defined as
\begin{eqnarray}
  \bm{f}(\bm{x}, \Omega t)= \bm{f}_u(\Omega t) A(\bm{x}), \quad
  \bm{f}_u(\Omega t)= \bm{f}_u(\Omega t+2\pi), \quad
  A(\bm{x})=
   \begin{cases}
     1 & (x_c - \frac{1}{2}l_x \le x \le x_c + \frac{1}{2}l_x, y_c - \frac{1}{2}l_y \le y \le y_c + \frac{1}{2}l_y)\\
     0 & (\textrm{otherwise})
   \end{cases}
   \label{eq:def of uniform force}
\end{eqnarray}

We highlighted the followings key points:
(1) In cases where the force is consistently parallel to a constant vector $\bm{f}_0$, we can describe $\bm{f}_u(t)$ as:
\begin{equation}
  \bm{f}_u(t) = \epsilon f_t(t)\bm{f}_0.
   \label{eq:def of temporal force}
\end{equation}
(2) The components of $\bm{F}(t)$ outside the rectangular area are all zero
 (c.f. Eq. (\ref{eq:relatioship betw. f_d and f})).
Consequently, the coupling function $\Gamma(\psi)$ is solely determined by the values of $\bm{q}$ within the specified rectangular area.

Finally, to address the optimization problems discussed in
 Sections \ref{sec: Case A: Minimum energy that enables lock-in phenomena} and
 \ref{sec: Case B: Maximum lock-in region of frequency} for the external force,
 as defined in Eq. (\ref{eq:def of uniform force}),
 Eqs. (\ref{eq:optimal external force for minimum energy}) and (\ref{eq:minimizer for the external force of maximum lock-in range})
 can be used by assuming that the components of $\bm{Z}$ corresponding to the components of $\bm{f}$ outside the rectangle to be absent.

These characteristics of the periodic uniform forcing within a spatially localized area facilitate the design and analysis of the optimal external force,
 as discussed in the subsequent sections.

\section{Result}
\subsection{Flow and phase sensitivity vector}
In this subsection, the flow details and the phase sensitivity vector field are described, and the optimal external forces for the lock-in phenomena are discussed.

\begin{figure}[h]
  \centering
  \includegraphics[width=0.90\textwidth]{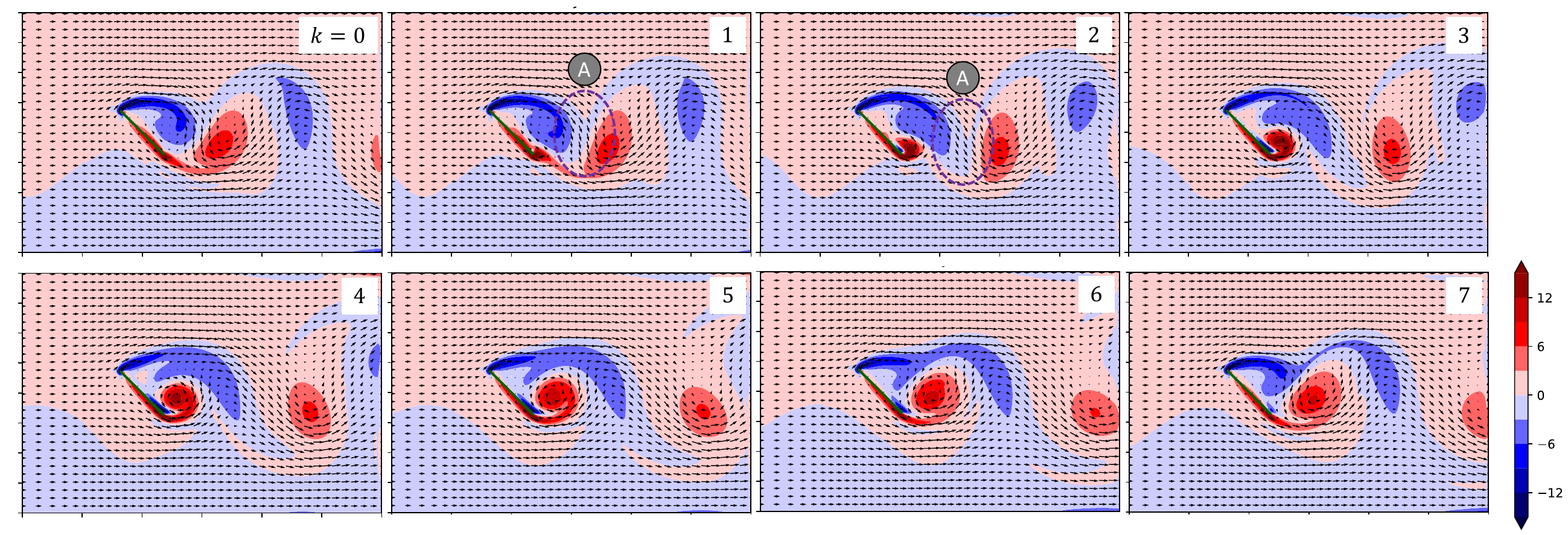}
  \caption{The velocity and vorticity fields. Shapshots at the phase $\phi/2\pi=k/8\;(k=0,1,\cdots,7)$ are shown.}
  \label{fig:Flow(aoa=45)}
\end{figure}
In the present condition, the flow converged to a periodic state.
The vorticity fields and the flow fields of the periodic solution are shown in Fig. \ref{fig:Flow(aoa=45)},
 where eight snapshots are shown with equal phase difference, $\phi/(2\pi)=k/8\;(k=0,1,\cdots,7)$.

Leading edge vortex (LEV) and trailing edge vortex (TEV) were generated periodically owing to the uniform flow and their interactions with the plate.
The LEV developed ($0 \le k \le 3$) and splits owing to TEV growth ($4 \le k \le 7$).
Part of the LEV remains for the redevelopment.

In contrast, TEV develops ($3 \le k \le 6$) to be swept by the flow induced by LEV ($k=7, 0$)
 to pinch off ($k=2$).
The volume (area) of the remaining TEV was not as large as that of the LEV, and the main body of the TEV developed near its trailing edge.
Thus, major vortex interactions occurred on the rear side of the plate.

\begin{figure}[h]
  \centering
  \includegraphics[width=0.90\textwidth]{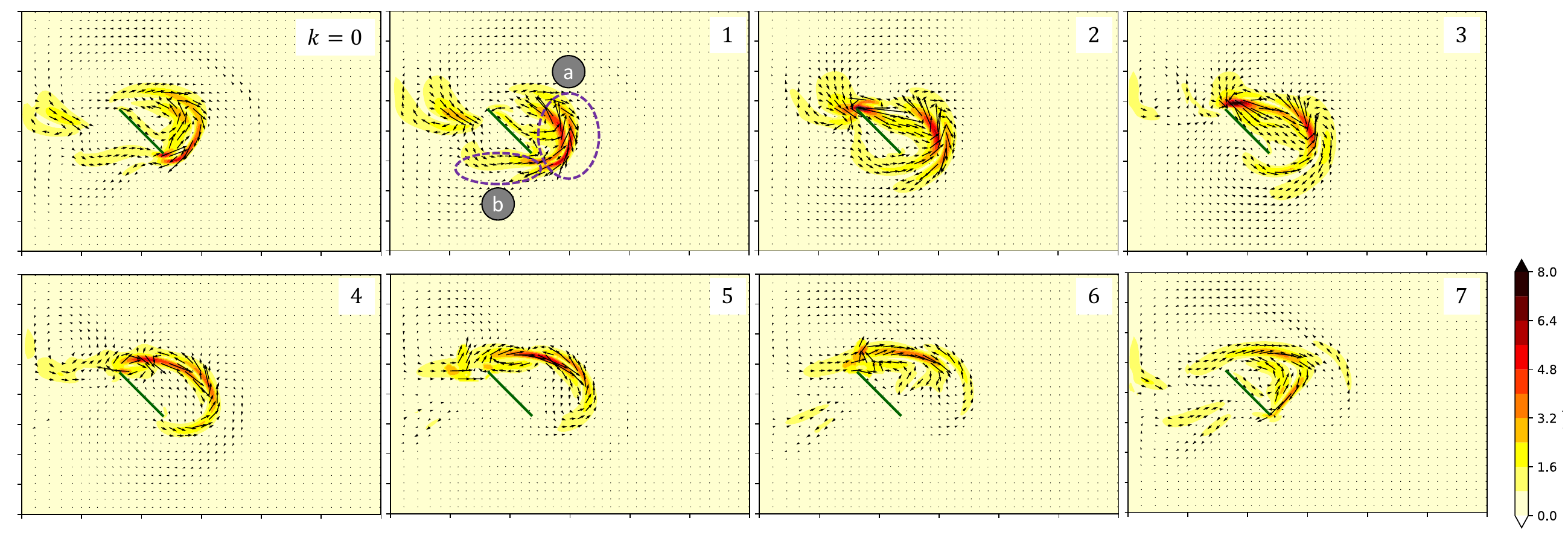}
  \caption{The fields of the phase sensitivity vector. The contour indicates the magnitude $|\bm{q}(\bm{x}, \phi)|$. Snapshots at the phase $\phi/2\pi=k/8\;(k=0,1,\cdots,7)$ are shown.}
  \label{fig:PSV(aoa=45)}
\end{figure}
Figure \ref{fig:PSV(aoa=45)} shows the phase sensitivity vector field $\bm{q}(\bm{x}, \phi)$.

The region exhibiting a pronounced phase response to the perturbation ($|\bm{q}(\bm{x}, \phi)|>1.6$) featured a distinctive spatial structure characterized by narrow, curve-like formations.
Although the specific configuration of these structures varies with the phases, a typical pattern on the backside of the plate comprised two nearly parallel curve-like structures (e.g. marked as ``a'' in Fig. \ref{fig:PSV(aoa=45)} ($k=1$)).
Furthermore, stronger response regions ($|\bm{q}(\bm{x}, \phi)|>3.2$) were primarily observed close to the leading edge, the trailing edge, and the region behind the plate where LEV and TEV interact.

When comparing the vortex dynamics with the structure of the phase response vector, a portion of the influence of $\bm{q}(\bm{x},\phi)$ can be attributed to the evolution of the flow, as outlined below, as outlined below.
However all aspects of the flow evolution are not captured by $\bm{q}(\bm{x}, \phi))$.
The vortex fields shown in Fig. \ref{fig:Flow(aoa=45)} ($k=1$) and ($k=2$) revealed certain features, and only the size of the LEV changed.
The TEV flowed downward and subsequently pinched off.
The flow field situated between LEV and TEV exhibited a negative $y-$direction, indicated by ``A'' in Fig. \ref{fig:Flow(aoa=45)} ($k=1$).
Region A shifted downstream, as shown in Fig. \ref{fig:Flow(aoa=45)} ($k=2$).
Furthermore, Fig. \ref{fig:PSV(aoa=45)} ($k=1)$ illustrates that the perturbations advance the phase.
The structure of the phase sensitivity vector in the negative $y$ flow region weakens the flow in the vicinity of the near-plate part of region A, while it strengthens the flow in the far-plate part.
This alternation encouraged a change in the flow behavior to that observed at $k=2$.

Another consequence of $\bm{q}(\bm{x},\phi)$ is a modification in the timing of the separation.
The earlier pinch-off of TEV during Fig. \ref{fig:Flow(aoa=45)} ($k=1$) and ($k=2$) occurs when TEV exhibits more rapid growth. 
This phenomenon is suggested by the structure of $\bm{q}(\bm{x},\phi)$ located upstream of the trailing edge, denoted by ``b'' in Fig.\ref{fig:PSV(aoa=45)} ($k=1)$. 
Unlike the previously mentioned double-curved structure mentioned previously, this region did not exhibit such features.

\subsection{Fourier spectrum of the phase sensitivity vector}
\label{sec: Fourier spectrum of the phase sensitivity vector}
In this subsection, the frequency decomposition of the phase sensitivity vector, used to examine the frequency-dependent characteristics, is discussed.
As each component of $\tilde{\bm{q}}(\bm{x};m)=(\tilde{q}_u(\bm{x};m), \tilde{q}_v(\bm{x};m))$ represents a field of the complex number,
 we displayed $\tilde{q}_j(\bm{x};m)\;(i=u,v)$ as a vector in the form of
$\displaystyle
  (\textrm{Re}(\tilde{q}_j(\bm{x};m)), \textrm{Im}(\tilde{q}_j(\bm{x};m)))
$.
In this presentation, the magnitude of the vector, denoted as $|\tilde{q}_j(\bm{x};m)|$, indicates
 the strength of the phase sensitivity to sinusoidal perturbations with angular frequency $m\Omega$ (c.f. Eq. (\ref{eq:counpling function by Z(m) and f_d(m)})).
The angle between the vector and the $x-$axis corresponds to $\textrm{arg}(\tilde{q}_j(\bm{x};m))$.
Notably, the area in which the argument of a complex number is uniform (vectors are parallel) signifies an area
 with a relatively pronounced phase response to uniform forcing.

\begin{figure}[h]
  \centering
  \includegraphics[width=0.9\textwidth]{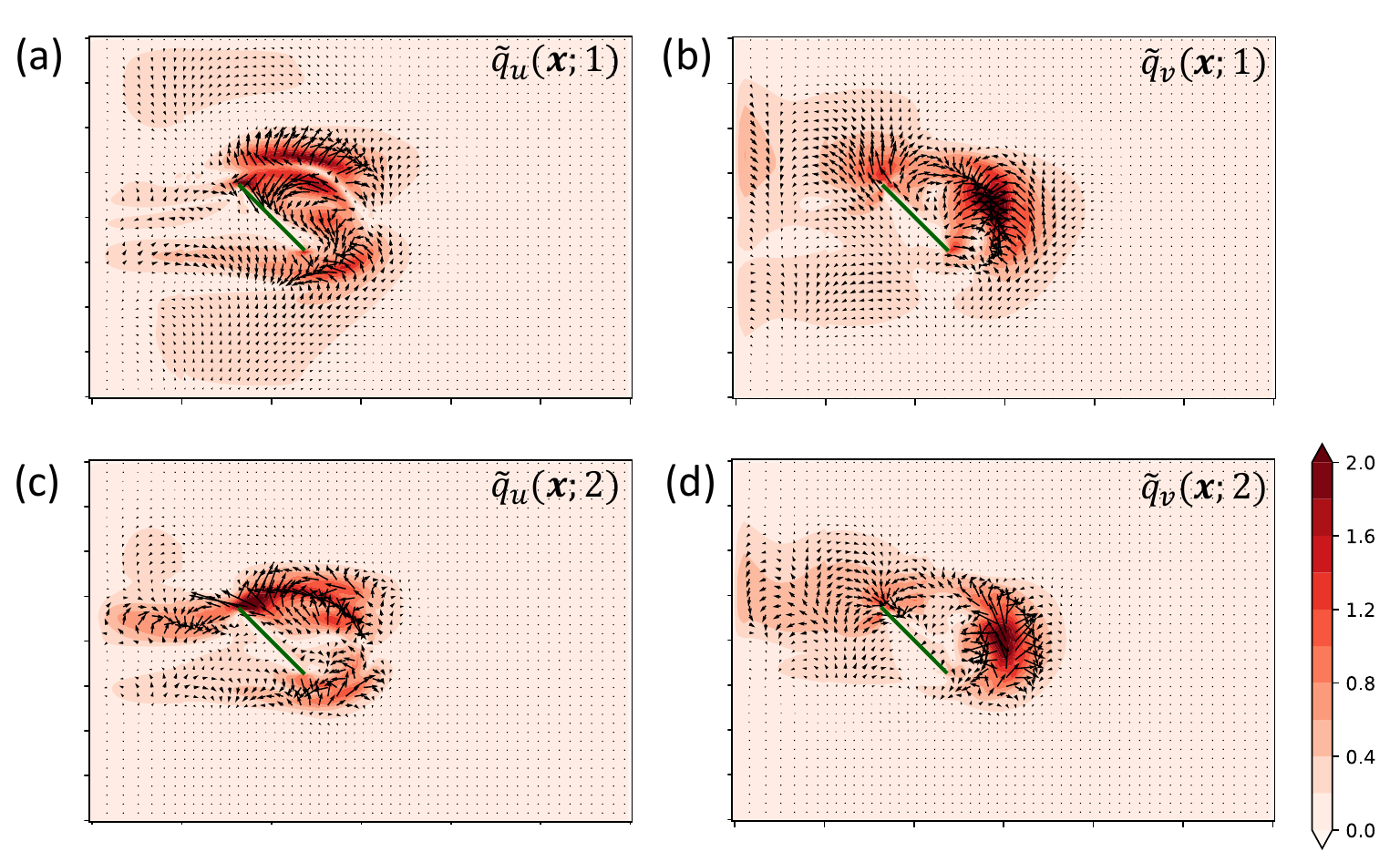}
  \caption{Phase responses to peroidic forcing, $\tilde{q}_u(\bm{x};m)$ and $\tilde{q}_u(\bm{x};m)$. Arrows and contours indicate the values (complex numbers in a Gauss plane) and their amplitudes, respectively. (a) $\tilde{q}_u(\bm{x};1)$, (b) $\tilde{q}_v(\bm{x};1)$, (c) $\tilde{q}_u(\bm{x};2)$ (d) $\tilde{q}_v(\bm{x};2)$.}
  \label{fig:Zuv12}
\end{figure}
The response to the perturbation at the angular frequency $\Omega$ is depicted in Fig.\ref{fig:Zuv12}(a) and (b).
The quantities $\tilde{q}_u(\bm{x};1)$ and $\tilde{q}_u(\bm{x};1)$ identify different regions that exhibit a strong response to the periodic perturbation.
Specifically, regions with large $|\tilde{q}_u(\bm{x};1)|$ are predominantly situated downstream of both the leading and trailing edges, whereas regions with large $|\tilde{q}_v(\bm{x};1)|$ are primarily found downstream of the middle of the plate.
In summary, the phases within these regions exhibited relatively minor variations,
 suggesting that a uniform periodic external force was effective when applied to each of these distinct areas.
However, notably, the specific phase values depended on the region, implying that the lock-in phase varies based on the location.

The response to a perturbation with an angular frequency $2\Omega$ is depicted in Fig.\ref{fig:Zuv12} (c) and (d).
The overall magnitude characteristics closely resembled those observed for $m=1$.
However, slight shifts were observed in the specific downstream regions,
 and the phase changed more rapidly within each area compared to the $m=1$ scenario.
Notably, a broader area upstream of the leading edge was observed where $|\tilde{q}_v(\bm{x};2)|$ assumed higher values
 ($|\tilde{q}_v(\bm{x};2)|>0.4$).
In Section \ref{sec: Lock-in details for the case C; the uniform periodic external forces within a rectangle region},
 the details of the response are discussed.

\begin{figure}[h]
  \centering
  \includegraphics[width=0.9\textwidth]{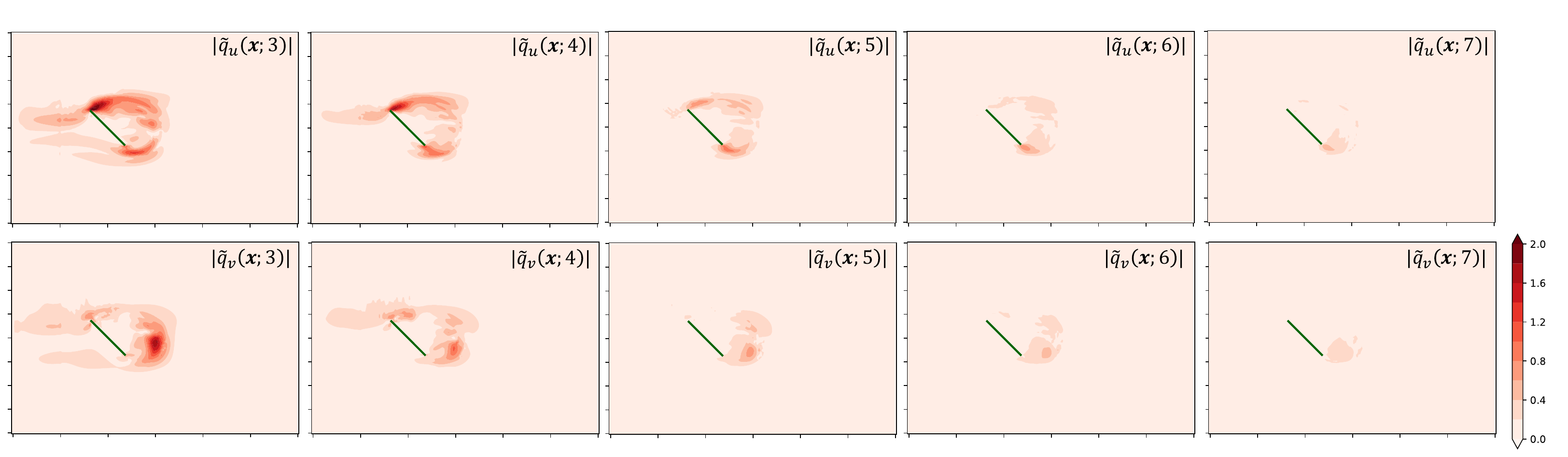}
  \caption{$|\tilde{q}_u(\bm{x};m)|, |\tilde{q}_v(\bm{x};m)| (3 \le m \le 7)$.}
  \label{fig:ZuZvamp3_7}
\end{figure}
The response to the perturbation with an angular frequency $m\Omega\;(m \ge 3)$ is depicted in Fig.\ref{fig:ZuZvamp3_7}.
As the value of $m$ increased, both $|\tilde{q}_u(\bm{x};m)|$ and $|\tilde{q}_v(\bm{x};m)|$ exhibited a decrease in magnitude.
Simultaneously, the widths of isophase lines for $\tilde{q}_u(\bm{x};m)$ and $\tilde{q}_v(\bm{x};m)$ diminished with increasing $m$,
 indicating a downstream advection of the flow structure.
Apparently for larger values of $m$, the region of strong response remained close to the trailing edge.

\subsection{Lock-in details for the case C: uniform periodic external forces in a rectangle region}
\label{sec: Lock-in details for the case C; the uniform periodic external forces within a rectangle region}
\subsubsection{Where is the best area for largest frequency range of the lock-in?}
\label{sec: Where is the best area for largest frequency range of the lock-in?}
We considered the frequency lock-in phenomenon induced by uniform periodic external forces within a rectangular region,
 as defined by Eqs.(\ref{eq:def of uniform force}) and (\ref{eq:def of temporal force}).
Specifically, we concentrated on the scenario where $f_t(t)=\sin mt$, and $\bm{f}_0=(1,0)$ and $(0,1)$, forming a basis of $\mathbb{R}^2$.
The sizes of the rectangular areas were selected as $(l_x, l_y)=(1.0, 0.5)$ and $(0.5, 1.0)$,
 which closely matched those of the regions where
 $|\tilde{q}_u(\bm{x};m)|$ or $|\tilde{q}_v(\bm{x};m)|$ exhibited significant values.
By varying the position vector $\bm{x}_c$, we derived the scalar field
 $\Gamma_{\textrm{max}}=\textrm{max}_{\psi \in [0,2\pi]} \Gamma(\psi)$ as a function of $\bm{x}_c$,
 indicating the extent of the frequency lock-in range.
 
Hereafter, the scalar field is denoted by ``$\Gamma_{\textrm{max},x}^{(m)}$ for $(l_x, l_y)=(1.0, 0.5)$''
 (the subscripts ``$x$'' and ``$y$'' of $\Gamma_{\textrm{max}}$ denote $\bm{f}_0=(1,0)$ and $\bm{f}_0=(1,0)$, respectively), for instance.

\begin{figure}[h]
  \centering
  \includegraphics[width=0.9\textwidth]{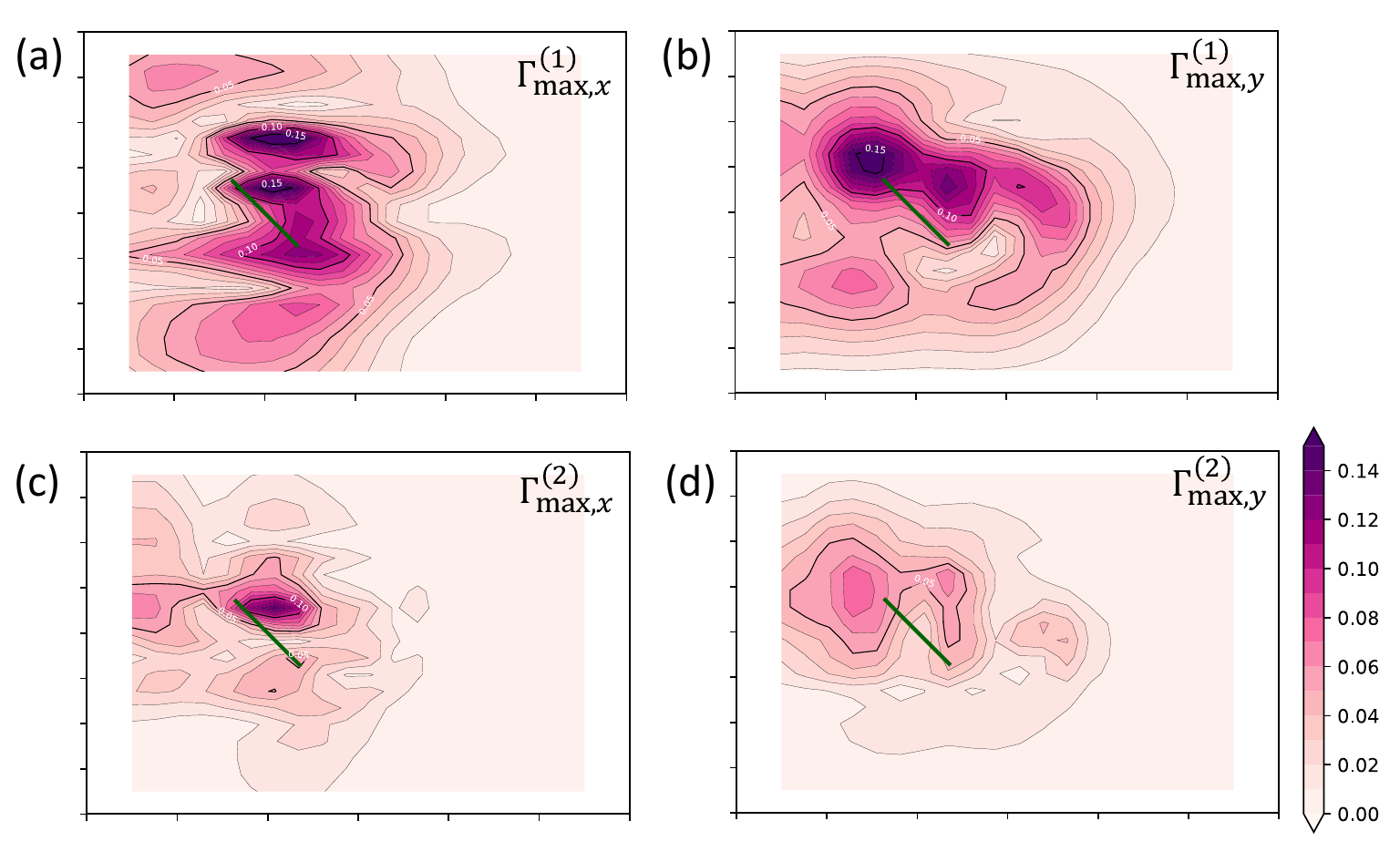}
  \caption{
(a) The field of $\Gamma_{\textrm{max},x}^{(1)}$ for $(l_x, l_y)=(1.0, 0.5)$ as function of $\bm{x}_c$.
(b) Same as (a), but for $\Gamma_{\textrm{max},y}^{(1)}$.
(c) Same as (a), but for $\Gamma_{\textrm{max},x}^{(2)}$.
(d) Same as (c), but for $\Gamma_{\textrm{max},y}^{(2)}$.}
  \label{fig:Gammaxy_1_0.5}
\end{figure}
Figures \ref{fig:Gammaxy_1_0.5}(a) and (b) depict $\Gamma_{\textrm{max}}$ for $(l_x, l_y)=(1.0, 0.5)$ and $m=1$
 for the external forces in the $x-$ and $y-$directions, denoted as $\Gamma_{\textrm{max},x}^{(1)}$ and $\Gamma_{\textrm{max},y}^{(1)}$, respectively.
The chosen rectangular dimensions of $1.0 \times 0.5$ roughly resembled the region where $|\tilde{q}_u(\bm{x};m)|$ was significant.
In this context, the field of $\Gamma_{\textrm{max},x}^{(1)}$ approximated $|\tilde{q}_u(\bm{x};m)|$ to some extents.
While the rectangle was not similar to the region with large $|\tilde{q}_v(\bm{x};m)|$,
 the maximum of $\Gamma_{\textrm{max},y}^{(1)}$ was comparable to $\Gamma_{\textrm{max},x}^{(1)}$
 since the argument of $|\tilde{q}_v(\bm{x};m)|$ exhibited more uniformity compared to that of $\tilde{q}_u(\bm{x};m)$.
The positions where $\Gamma_{\textrm{max},x}^{(1)}$ and $\Gamma_{\textrm{max},y}^{(1)}$ attain their maximum values are detailed in Table \ref{table:Gamma_max}, demonstrating that these maximum values to be comparable.

Figures \ref{fig:Gammaxy_1_0.5} (c) and (d) depict $\Gamma_{\textrm{max}}$ for $(l_x, l_y)=(1.0, 0.5)$ and $m=2$.
As the phase changed rapidly within regions where $|\tilde{q}_u(\bm{x};m)|$ or $|\tilde{q}_v(\bm{x};m)|$
 assumed larger values,
 $\Gamma_{\textrm{max},x}^{(2)}$ exhibited smaller values compared to $|\tilde{q}_u(\bm{x};m)|$
 with a few exceptional regions in the downstream and the upstream of the leading edge.
Similar characteristics were observed for the $\Gamma_{\textrm{max},y}^{(2)}$ field.
While the maximum value of $\Gamma_{\textrm{max},x}^{(2)}$ was relatively higher compared to other instances $m=2$, the peak did not exhibit broader support.

\begin{figure}[h]
  \centering
  \includegraphics[width=0.9\textwidth]{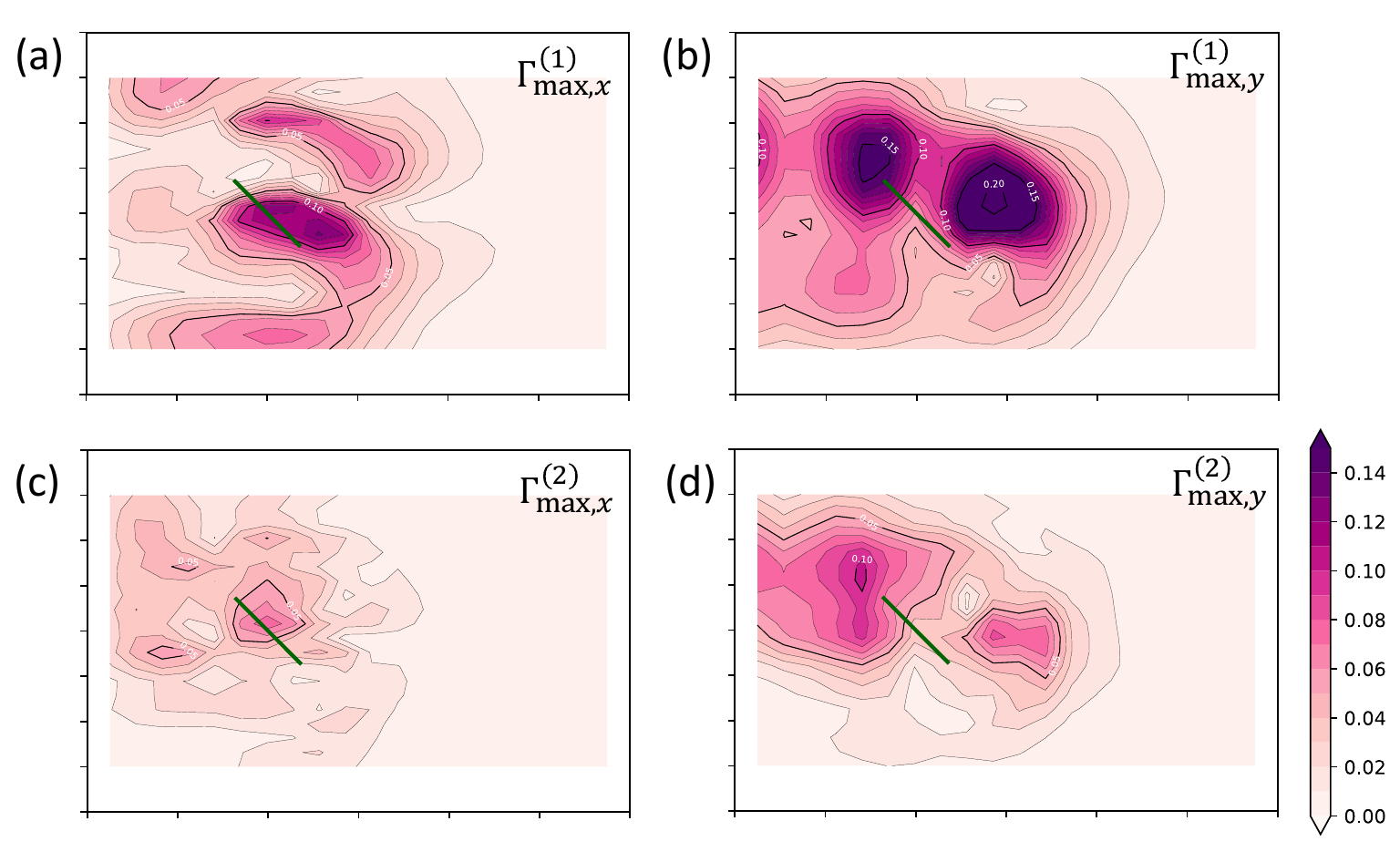}
  \caption{Same as Fig. \ref{fig:Gammaxy_1_0.5}, but for $(l_x, l_y)=(0.5, 1.0)$.}
  \label{fig:Gammaxy_0.5_1}
\end{figure}
Figures \ref{fig:Gammaxy_0.5_1} (a) and (b) depict $\Gamma_{\textrm{max}}$ for $(l_x, l_y)=(0.5, 1.0)$ and $m=1$.
Since the sizes of the rectangle, $0.5 \times 1.0$, closely resembled the region where $|\tilde{q}_v(\bm{x};m)|$ was significant,
 the $\Gamma_{\textrm{max},y}^{(1)}$ field approximated $|q_v(\bm{x};m)|$ to some extents.
However, in this case, the rectangle did not resemble a region with significant $|q_u(\bm{x};m)|$ values.
Notably, the maximum value of $\Gamma_{\textrm{max},y}^{(1)}$($\simeq 0.21$) exceeded the maximum value of  $\Gamma_{\textrm{max},x}^{(1)}$($\simeq 0.13$) (Table \ref{table:Gamma_max}).

Figures \ref{fig:Gammaxy_0.5_1} (c) and (d) depict $\Gamma_{\textrm{max}}$ for $(l_x, l_y)=(0.5, 1.0)$ and $m=2$.
Although the $\Gamma_{\textrm{max},x}^{(2)}$ field is generally weak,
 the $\Gamma_{\textrm{max},y}^{(2)}$ field exhibited a prominent peak in the upstream section of the leading edge.
The maximum value of $\Gamma_{\textrm{max},y}^{(2)}$($\simeq 0.10$) was comparable to
 the maximum value of $\Gamma_{\textrm{max},x}^{(2)}$ for $(l_x, l_y)=(1.0, 0.5)$ (about $0.14$).
However, the strong region' extent was broader.
The broad lock-in region in the upstream of the plate is characteristic of $\Gamma_{\textrm{max},y}^{(2)}$
 (for both $(l_x, l_y)=(1.0,0.5)$ and $(0.5,1.0)$).

\begin{table}[h]
\centering
\begin{tabular}{c|c|cc|c}
 mode & $(l_x, l_y)$ & $x^*$ &  $y^*$ &  Maximum value  \\
  \hline \hline
$\Gamma_{\textrm{max},x}^{(1)}$ & $(1.0, 0.5)$  &  2.079 & 2.829 & 0.1794\\
$\Gamma_{\textrm{max},y}^{(1)}$ & $(1.0, 0.5)$  &  1.553 & 2.645 & 0.1587\\
$\Gamma_{\textrm{max},x}^{(1)}$ & $(0.5, 1.0)$  &  2.566 & 1.763 & 0.1321\\
$\Gamma_{\textrm{max},y}^{(1)}$ & $(0.5, 1.0)$  &  2.855 & 2.079 & 0.2142\\
$\Gamma_{\textrm{max},x}^{(2)}$ & $(1.0, 0.5)$  &  2.079 & 2.276 & 0.1420\\
$\Gamma_{\textrm{max},y}^{(2)}$ & $(1.0, 0.5)$  &  1.289 & 2.645 & 0.07749\\
$\Gamma_{\textrm{max},x}^{(2)}$ & $(0.5, 1.0)$  &  1.987 & 2.079 & 0.07769\\
$\Gamma_{\textrm{max},y}^{(2)}$ & $(0.5, 1.0)$  &  1.408 & 2.553 & 0.1045\\
\end{tabular}
\caption{The position of rectangle $\bm{x}^*=(x^*, y^*)$ where maximum frequency lock-in is obtained and the maximum value of $\Gamma_{\textrm{max}}$.}
\label{table:Gamma_max}
\end{table}

\subsubsection{Arnold's tongue: Two cases}
We examined two cases: Cases I and II.
 (I) Maximizing the lock-in frequency range. We considered $\Gamma_{\textrm{max},y}^{(1)}$ with $(l_x, l_y)=(0.5,1.0)$. We obtained $\Gamma_{\textrm{max}}=0.2142$ where $(x^*, y^*)=(2.855, 2.079)$ (Table \ref{table:Gamma_max}).
The rectangular region was located downstream of the plate.
 (II) Lock-in in the upstream region using an external force. We focused on $\Gamma_{\textrm{max},y}^{(2)}$ with $(l_x, l_y)=(1.0, 0.5)$. We obtained $\Gamma_{\textrm{max}}=0.1045$ where $(x^*, y^*)=(1.408, 2.553)$ (Table \ref{table:Gamma_max}).
Case II corresponded to a 2:1 frequency ratio lock-in induced by an external force. 
In addition, the forcing region (rectangle) was situated in the upstream of the leading edge,
 indicating that the perturbation was advected toward the leading edge, leading to entrainment.
A similar large response region in the upstream was reported in the case of the K\'arm\'an's vortex street
\cite{iima19_jacob_free_algor_to_calcul}.

To determine the lock-in region,
 we conducted direct numerical simulations with periodic external forces for
 either 500 periods ($\epsilon \ge 0.05$)
 or 1000 periods ($\epsilon < 0.05$).
The initial conditions for these simulations were periodic solutions in the absence of external forces.
The period of the system under the influence of an external force was determined by the peak-to-peak duration
 of the lift coefficient.
We calculated the average of the last 100 periods to estimate the period and the standard deviation to assess whether frequency lock-in occurred or not.
The angular frequency of the system under the influence of the external force, as obtained through numerical simulation, is denoted by $\omega_{\textrm{sim}}$.

\begin{figure}[h]
  \centering
  \includegraphics[width=0.9\textwidth]{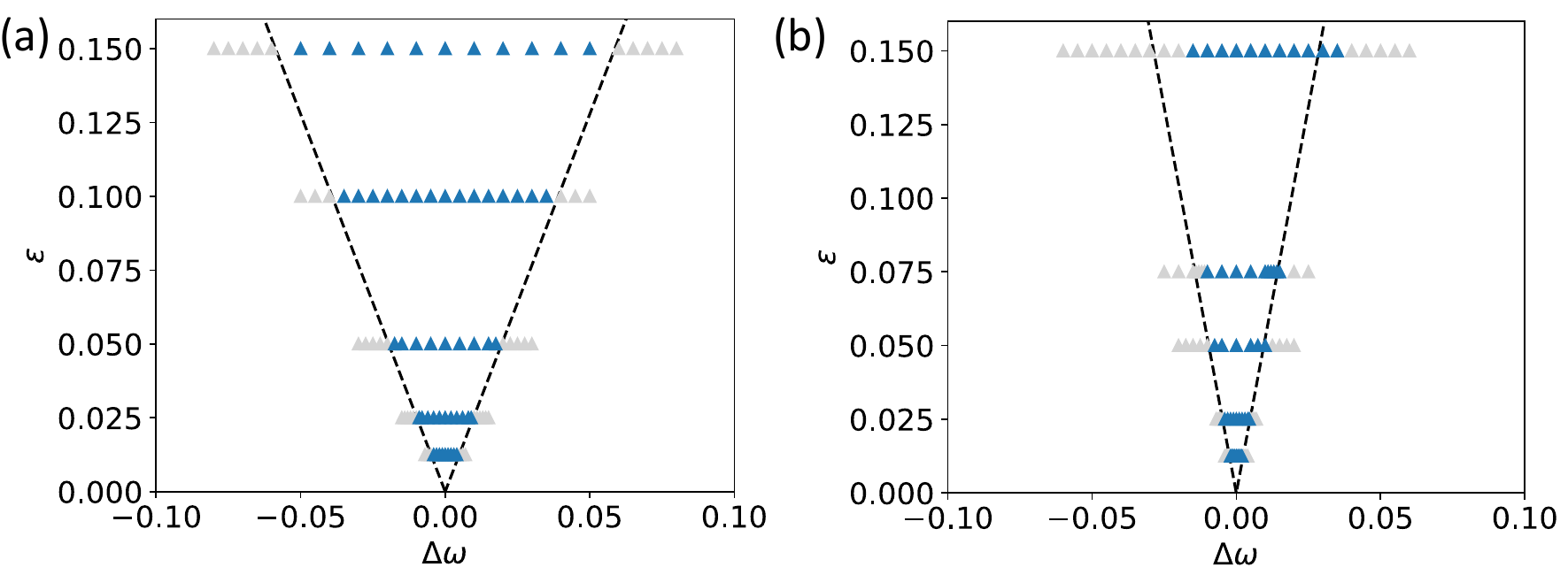}
  \caption{Arnold's tongues. Broken lines indicates theoretical prediction of the lock-in. Blue triangles and gray triangles indicate the lock-in state and no lock-in states, respectively. (a) The case I. (b) The case II.}
  \label{fig:Arnold}
\end{figure}

\begin{figure}[h]
  \centering
  \includegraphics[width=0.9\textwidth]{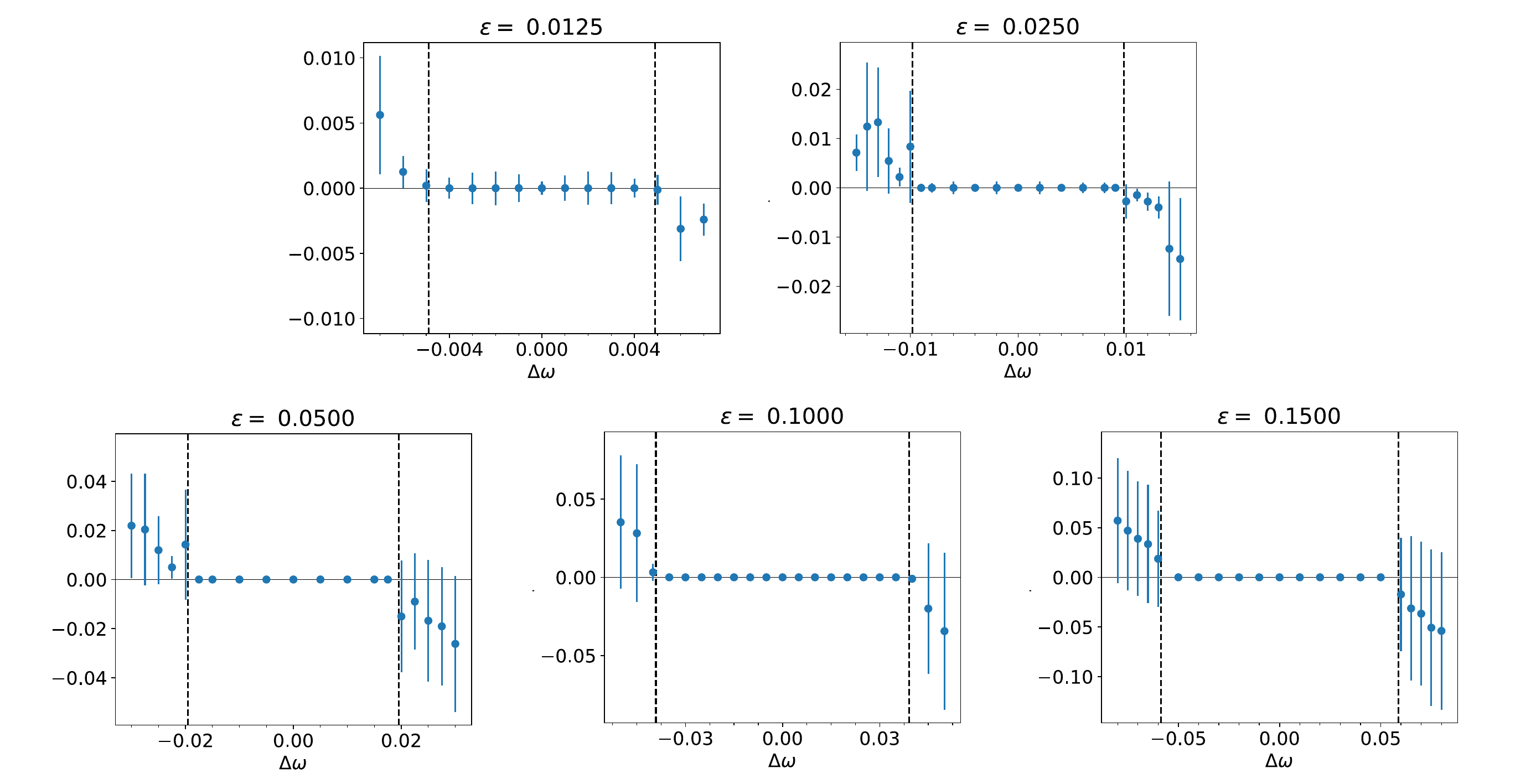}
  \caption{Difference between the angular frequency of the external force and that observed in the simulation, $\Omega - \omega_{\textrm{sim}}$, for the case I. Vertical broken lines indicate the theoretical boundary of the lock-in region.}
  \label{fig:angular frequency difference: case I}
\end{figure}

Figure \ref{fig:Arnold}(a) shows the lock-in region for Case I in $\Delta \omega$-$\epsilon$ plane.
The blue triangles represent instances where the frequency lock-in was observed, with the condition:
 $|\Omega - \omega_{\textrm{sim}}| < \epsilon_{e}$
 where
 $\epsilon_{e}=1.0\times 10^{-4}$.
Conversely, the gray triangles indicate cases where the frequency lock-in was not observed.
The dashed lines delineate the boundary of the lock-in region, as predicted by the phase reduction theory (see Eq. (\ref{eq:lock-in condition}))
 over the entire investigations of $0 \le \epsilon \le 0.15$.
The predictions of the phase reduction theory align closely with the results from direct numerical simulations.
To examine the details of the lock-in phenomena, we displayed the values of $\Omega - \omega_{\textrm{sim}}$ in Fig. \ref{fig:angular frequency difference: case I} for each $\epsilon$.
This representation also highlights the good agreement between the theoretical predictions and the simulation results.

\begin{figure}[h]
  \centering
  \includegraphics[width=0.9\textwidth]{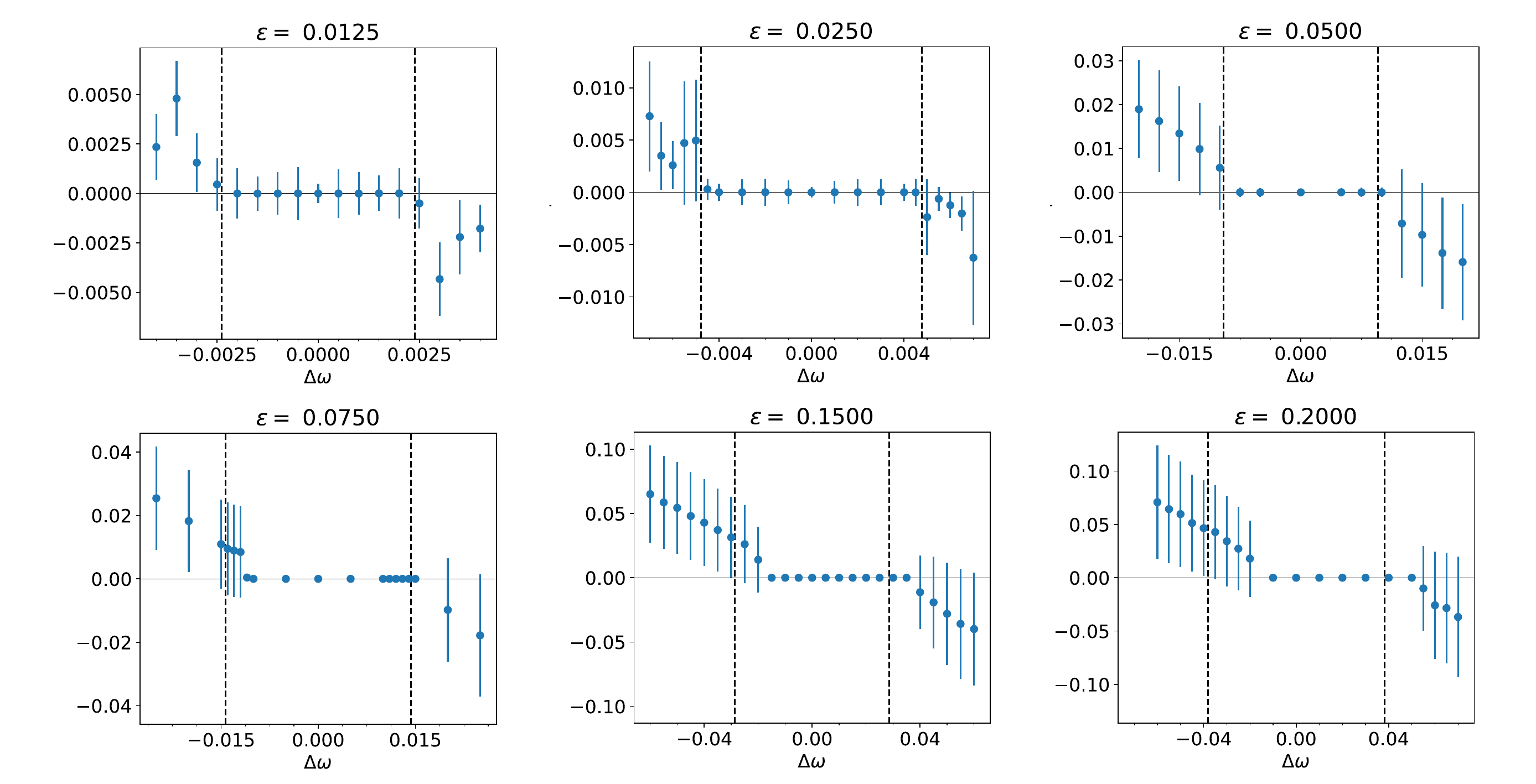}
  \caption{Same as Fig. \ref{fig:angular frequency difference: case I}, but for the case II.}
  \label{fig:angular frequency difference: case II}
\end{figure}

Figure \ref{fig:Arnold}(b) shows the lock-in region for Case II in $\Delta \omega$-$\epsilon$ plane.
In this case, the theoretical predictions closely matched the numerical results when $\epsilon \le 0.05$; however they deviated from each other when $\epsilon > 0.05$.
The lock-in range for the region where $\epsilon > 0.05$ shifted to the higher frequency side as $\epsilon$ increased.
Examining the details of the lock-in phenomena, we displayed the values of $\Omega - \omega_{\textrm{sim}}$ in Fig. \ref{fig:angular frequency difference: case II} for each $\epsilon$.

\subsection{Optimal external forces that achieves the frequency lock-in and comparison with uniform force within the rectangle region}
\label{sec:Optimal external forces and its efficiency}
In this subsection, the optimal external forces in following two distinct cases are discussed:
(A) minimizing the energy of the external force to achieve frequency lock-in, and
(B) maximizing the frequency range for lock-in with a constant energy external force.
These cases were previously introduced in Section \ref{sec:Optimal external forces under several conditions}.

\subsubsection{Coupling functions}
\begin{figure}[h]
  \centering
  \includegraphics[width=0.9\textwidth]{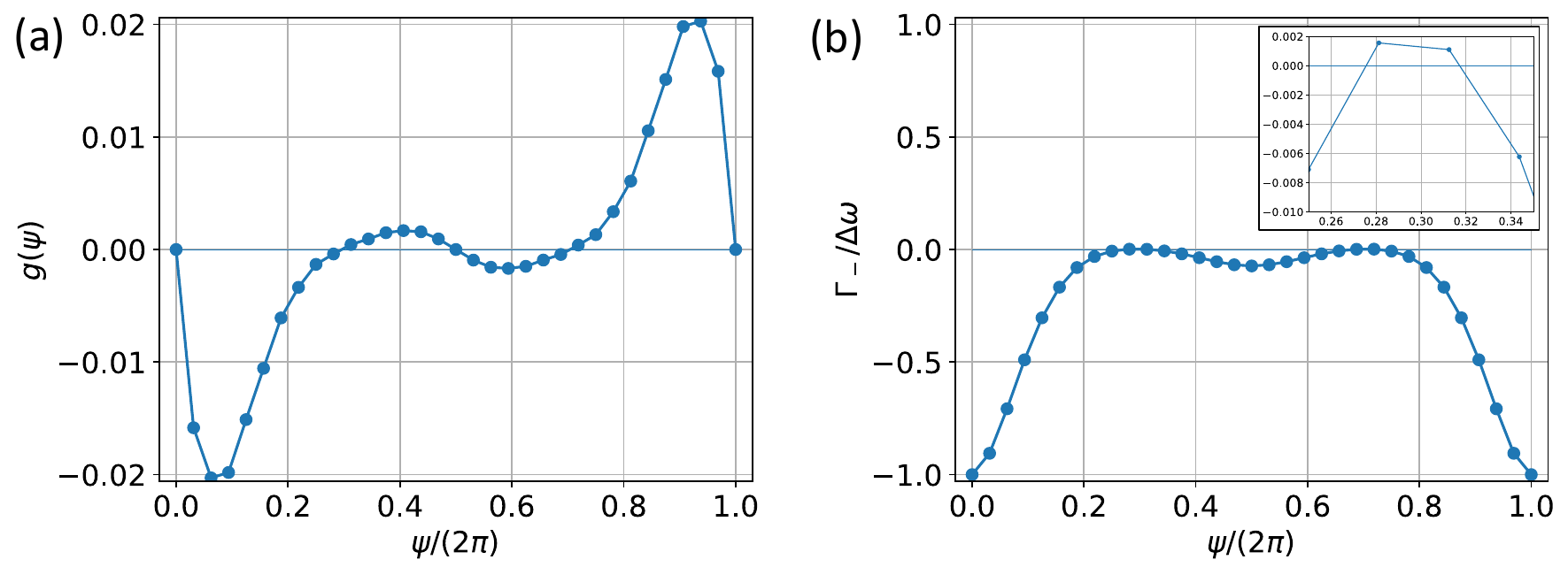}
  \caption{(a) $g(\psi)$. (b) $\Gamma_-(\psi)$.}
  \label{fig:function g, gamma_psi-}
\end{figure}
Regarding Case A, the shapes of the coupling function (\ref{eq: Case A: coupling function}) for $\psi_{\pm}$ were identical
 with the only difference of the phase.
The phase difference $\Delta \psi$ can be determined from Eq.(\ref{eq:Delta psi}).
Assuming $\psi_{-}=0$(Eq.(\ref{eq:property of g(x)})), one of the solutions is $\Delta \psi=\psi_{+}=\pi$.

Figure \ref{fig:function g, gamma_psi-} (a) shows the function $g(\psi)$.
Four solutions were obtained within the region $[0,2\pi)$, including one nontrivial solution $\psi^*/(2\pi) =0.2980$ ($\psi^* = 1.0249$).
Notably, $2\pi-\psi^*$ also satisfies the equation according to Eq. (\ref{eq:property of g(x)}).
Therefore, the complete set of solutions includes: $\psi=0, \psi^*, \pi$ and $2\pi-\psi^*$.

Figure \ref{fig:function g, gamma_psi-} (b) shows the coupling function for $\psi_{-}(=0)$ in a non-dimensional form by $\Delta \omega$.
As expected, the lowest angular frequency for lock-in was achieved at $\psi=\psi_{-}$.
For smaller frequency differences, four lock-in phases were encountered.

\begin{figure}[h]
  \centering
  \includegraphics[width=0.9\textwidth]{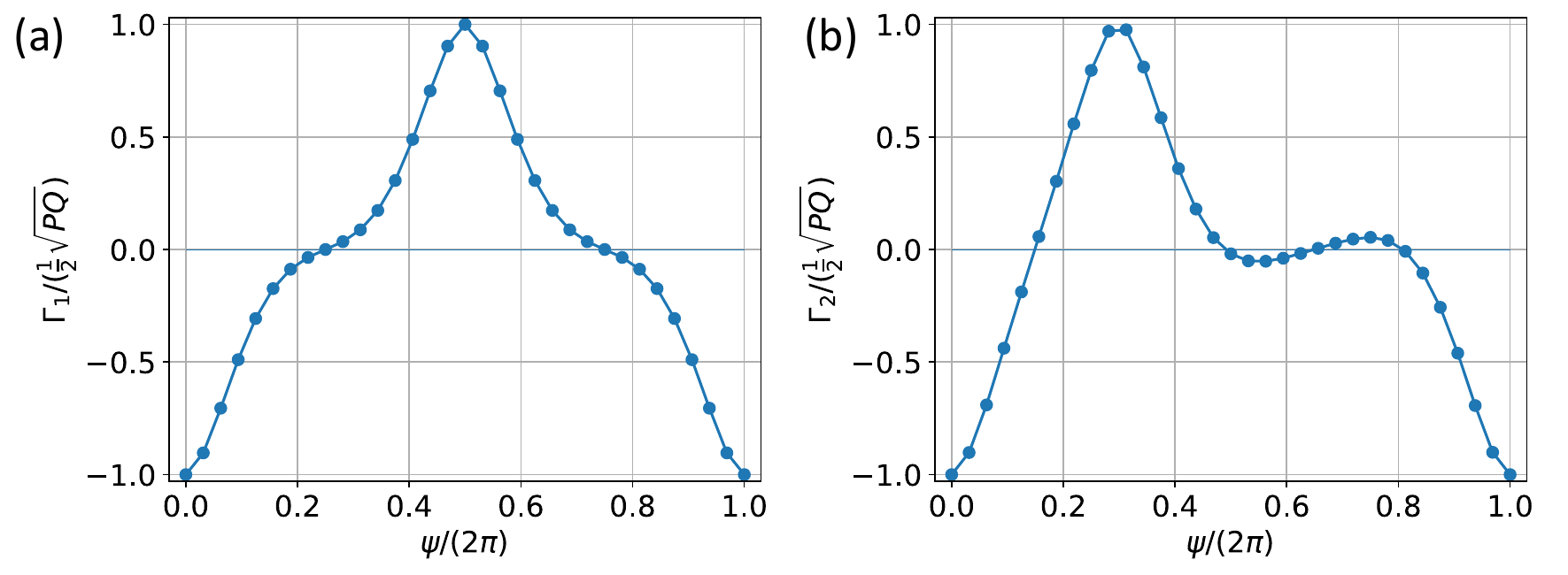}
  \caption{(a) $\Gamma_1(\psi)$ ($\psi_+=\pi$). (b) $\Gamma_2(\psi)$ ($\psi_+=\psi^*$).}
  \label{fig:largestEntrainment gamma}
\end{figure}

Regarding Case B, the coupling functions defined in Eq. (\ref{eq: Case B: coupling function}) were obtained from the solution of Eq. (\ref{eq:Delta psi}).
We consider two specific cases $\psi_+=\pi$, denoted by $\Gamma_1$, and $\psi_+=\psi^*$, denoted by $\Gamma_2$.
These choices yielded distinct coupling functions, which are illustrated in Figs.\ref{fig:largestEntrainment gamma}(a) and (b), and we displayed in a non-dimensionalized form by $\frac12 \sqrt{PQ}$.

Evidently, as shown in Fig.\ref{fig:largestEntrainment gamma}(a), the shape of $\Gamma_1(\psi)$ exhibited symmetry with respect to the line $\psi=\pi$.
Moreover, the shapes within the ranges $0 \le \psi \le \pi$ and  the shape in $\pi \le \psi \le 2\pi$
 are anti-symmetric with respect to the points $(\psi, \Gamma_1)=(\pi/2,0)$ and $(\psi,\Gamma_1)=(3\pi/2,0)$, respectively.

These characteristics can be derived from the definition of $\Gamma_1$.
In fact, the following two identities hold true:
$\Gamma_1(\psi) = \Gamma_1(2\pi-\psi)$ and $\Gamma_1(\psi) = -\Gamma_1(\psi+\pi)$.
The first identity can be expressed as follows.
The definition of $\Gamma_1$ (Eq. (\ref{eq: Case B: coupling function})) with $\psi_+=\pi$ and $\psi_-=0$ implies that
 $\Gamma_1'(\psi)=\sqrt{\frac{P}{Q}}[g(\psi-\pi)-g(\psi)]$.
Then, $\Gamma_1'(\psi)=-\Gamma_1'(-\psi)$ holds due to the properties of $g(\psi)$ (Eq. (\ref{eq:property of g(x)})).
Integrating the relationship, we obtain $\Gamma_1(\psi)=\Gamma_1(2\pi-\psi)$.
The second identity can be easily shown through the definition of $\Gamma_1$.
These two identities indicate that $\Gamma_1(\pi-\psi)=\Gamma_1(2\pi-(\psi+\pi))=\Gamma_1(\psi+\pi)=-\Gamma_1(\psi)$.
Consequently, the following identities hold:
\begin{equation}
  \Gamma_1(\psi)=\Gamma_1(2\pi-\psi), \quad \Gamma_1(\pi-\psi)=-\Gamma_1(\psi), \quad \Gamma_1(2\pi-\psi)=-\Gamma_1(\pi+\psi),  
    \label{eq:properties of Gamma_1}
\end{equation}
 which correspond to the characteristics of $\Gamma_1$ as shown in Fig. \ref{fig:largestEntrainment gamma}(a).

Based on the properties (\ref{eq:properties of Gamma_1}), the lock-in phases for $\Delta \omega=0$ were
 $\psi/(2\pi)=0.25 (\textrm{unstable})$ and $0.75 (\textrm{stable})$.
The shape of the graph suggests absence of other lock-in phases.

As shown in Fig.\ref{fig:largestEntrainment gamma}(b), the symmetries displayed in Fig.\ref{fig:largestEntrainment gamma}(a) were not observed because $\psi_+ \ne \pi$.
The shape of $\Gamma_2(\psi)$ can yield multiple lock-in phases for smaller frequency difference within lock-in region.
The stable lock-in phases for $\Delta \omega=0$ were estimated as $\psi/(2\pi)= 0.491649 $ and $0.807218$ through the linear interpolation.

\subsubsection{Comparison between the uniform force within the rectangular region and optimal forces}
\label{sec: Comparison with the uniform force within the rectangular region}
In this section, the efficiencies of the uniform force within the rectangular region are discussed by comparing them with the optimal forces.
We focus on Case 1, where $\Gamma_{\textrm{max},y}^{(1)}$ for $(l_x, l_y)=(0.5, 1.0)$ (referred to as `the case of uniform force' hereafter), and examine the lock-in characteristics in comparison with the optimal forces calculated within the same rectangle region.
As previously discussed in Section \ref{sec: Case A: Minimum energy that enables lock-in phenomena},
 the optimal external force is proportional to $-\bm{Z}(\psi)$ (Eq.(\ref{eq:optimal external force for minimum energy})).
In the case where the external force was applied only within the rectangular region,
 the optimal force was determined using a procedure similar to that described in Section \ref{sec: Case A: Minimum energy that enables lock-in phenomena}.
The optimal external force is given by the Eq.(\ref{eq:optimal external force for minimum energy}) within a localized region.

\paragraph{Energy ratio with the same lock-in range}
The energy of the external force,
 $E_{\pm, \textrm{opt}}$,
 was obtained from the square of Eq.(\ref{eq:optimal external force for minimum energy}) as
$
  E_{\pm, \textrm{opt}} =
  \langle \left|\bm{F}_{\pm}\right|^2 \rangle = \frac{\Delta \omega_{\pm}^2}{\langle Z^2 \rangle},
$
 where $\Delta \omega_{\pm}$ corresponds to the cases of $\psi_{\pm}$, respectively.
Notably, $\psi_{+}/(2\pi)= 0.2606$ when the optimal force was calculated within the rectangle region.

In the case of uniform force, the energy of the external force, $E_{u}$, was:
$
  E_{u} = \langle \left|\bm{F}_{\textrm{u}}\right|^2 \rangle = \frac{\epsilon^2 M}{2},
$
 where $M$ is the number of grid points within the rectangle.
The value of $\langle Z^2 \rangle$ is $0.002306$ and $M=450$.
Further, as listed in Table \ref{table:Gamma_max}, $\Gamma_{\textrm{max}}=0.2142$.

The comparison under the condition that the maximum frequency differences for the lock-in, i.e.,
 $\Delta \omega = \epsilon \Gamma_{\textrm{max}}$,
 we obtain
\begin{eqnarray}
  \frac{E_{\pm, \textrm{opt}}}{E_{u}}  &=& \frac{2 \Gamma_{\textrm{max}}^2}{M \langle Z^2 \rangle} = 0.0884.\label{105553_16Oct23}
\end{eqnarray}
Thus, the energy of external force in the case of uniform force was approximately 8.8\% of the optimal energy. 
Notably, Eq. (\ref{105553_16Oct23}) does not depend on $M$ because $\Gamma \sim M^0$ and $Z \sim M^{-1}$ (c.f. Section.\ref{sec:Phase reduction theory}).

\paragraph{Lock-in range ratio with the same energy}
Similar calculation to obtain Figure \ref{fig:largestEntrainment gamma} provides that
 the maximum frequency difference of the optimal force within the rectangle region,
 $\Delta \omega_{\textrm{opt}}$,
 is $\Delta \omega_{\textrm{opt.}}= \frac12 \sqrt{P_{\textrm{opt}} Q_{\textrm opt}}$,
 where $P_{\textrm{opt}}$ is the energy of the optimal external force within the rectangular region,
 and $Q_{\textrm{opt}}(=0.004854)$ has the same definition as $Q$ but is calculated within the rectangular region.
In the case of uniform force, the maximum frequency difference for the lock-in,
 $\Delta \omega_{\textrm{rect}}$, 
 is $\epsilon \Gamma_{\textrm{max}}$.

Presumably, the energy of the case of uniform force,
 $E_u$,
 was the same as that of the optimal case, $P_{\textrm{opt}}$. i.e.,
$\displaystyle P_{\textrm{opt}}= \frac{M \epsilon^2}{2}$.
Then, the ratio of the maximum frequencies is:
\begin{equation}
  \frac{\Delta \omega_{\textrm{opt}}}{\Delta \omega_{\textrm{rect}}}
    = \frac{0.500 \sqrt{M Q_{\textrm{opt}}}}{\sqrt{2} \Gamma_{\textrm{max}}}
    = 2.44.
\end{equation}
Thus, the maximum lock-in range for the optimal force was 2.44 times that of the maximum lock-in range for the uniform force.

\paragraph{Field of optimal forces}
\begin{figure}[h]
  \centering
  \includegraphics[width=0.9\textwidth]{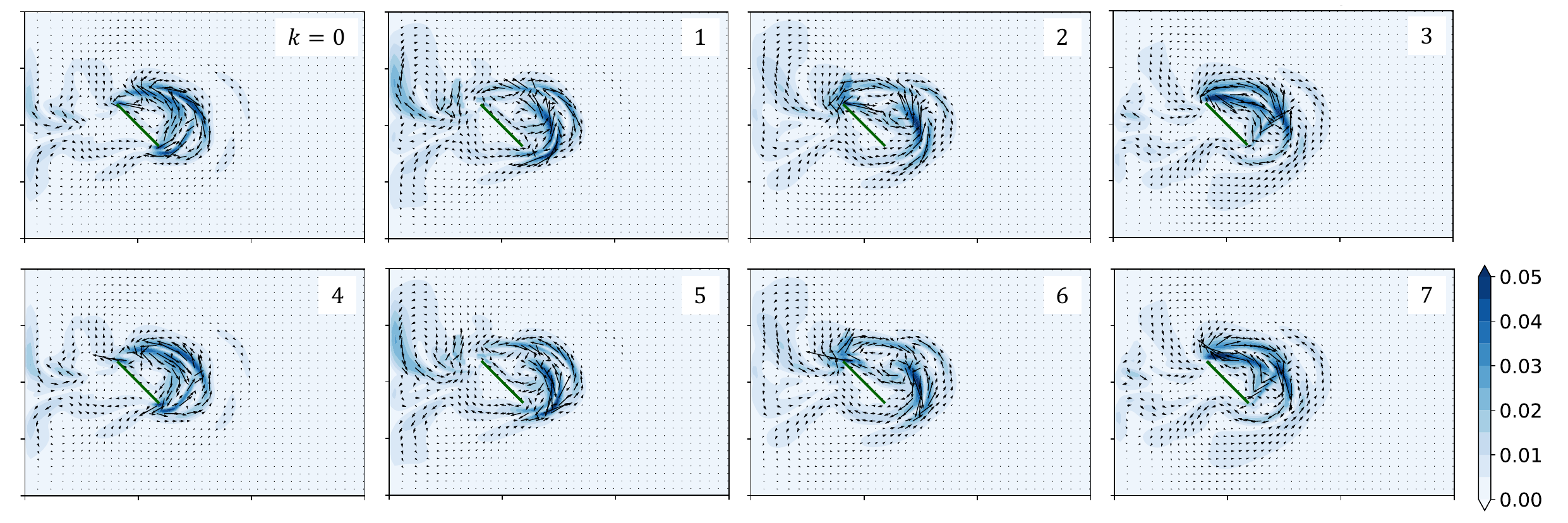}
  \caption{$\bm{f}_{*} (\psi_+=\pi)$}
  \label{fig:F* (psi+=pi)}
\end{figure}
Figure \ref{fig:F* (psi+=pi)} shows the fields of the optimal force of the maximum lock-in region,
 denoted as $\bm{f}_{*}(\bm{x}, \theta)$, in the case $\psi_+=\pi$.
The optimal force was converted to the force field using a similar formula as Eq. (\ref{eq:relationship betw. Z and Q}).
Notably, the force $\bm{f}_{*}(\bm{x}, \theta)$ satisfies the following relationship:
\begin{equation}
  \bm{f}_{*}(\bm{x}, \theta) = -\bm{f}_{*}(\bm{x}, \theta+\pi).
    \label{eq:symmetry of fe*}
\end{equation}
This relationship is derived from the definition (\ref{eq: Case B: coupling function}) and the settings of $(\psi_+, \psi_-)=(0, \pi) $.
In all the phases, the region with the strongest optimal force was predominantly situated behind the plate.
In several phases, such as $k=1,2$ (and $5,6$), a pronounced region of strong forces was observed upstream of the leading edge,
 suggesting a more favorable position and timing for the phase control in the upstream region.

\begin{figure}[h]
  \centering
  \includegraphics[width=0.9\textwidth]{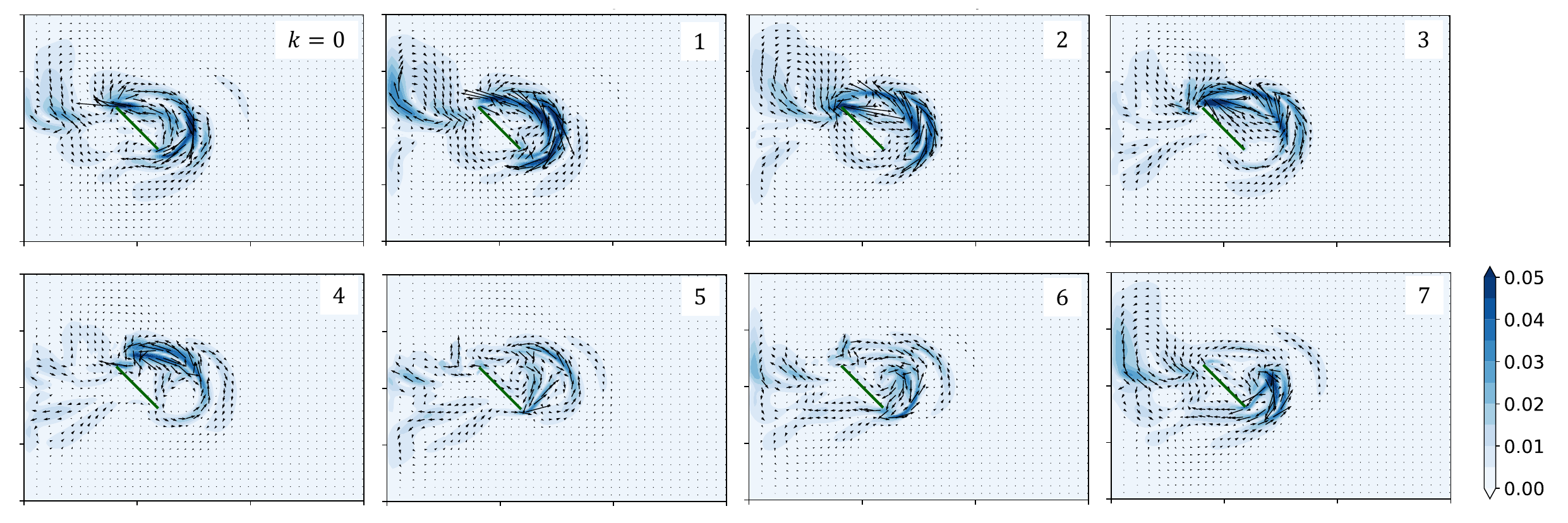}
  \caption{$\bm{f}_{*} (\psi_+=\psi^*)$}
  \label{fig:F* (psi+=psi*)}
\end{figure}
Figure \ref{fig:F* (psi+=psi*)} shows $\bm{f}_{*}(\bm{x}, \theta)$ in the case $\psi_+=\psi^*$.
An area with a strong force $\bm{f}_{*}(\bm{x}, \theta)$ was noticeable when $k$ was between $0$ and $3$,
 rather than when it was between $4$ and $7$.
Interestingly, the optimal forces for $\psi_{+}=\pi$ and $\psi^*$ differed, even though the maximum frequency lock-in region was nearly identical.
As shown in Fig. \ref{fig:F* (psi+=pi)}, the region with stronger forces was primarily found downstream of the plate,
 with a noticeable difference between the downstream of the leading edge and that of the trailing edge.
This implies that the forces mainly control the interaction between the LEV and TEV to achieve a frequency lock-in.
As shown in Fig. \ref{fig:F* (psi+=psi*)}, a region with stronger forces was observed downstream of the plate and upstream of the leading edge.

\begin{figure}[h]
  \centering
  \includegraphics[width=0.9\textwidth]{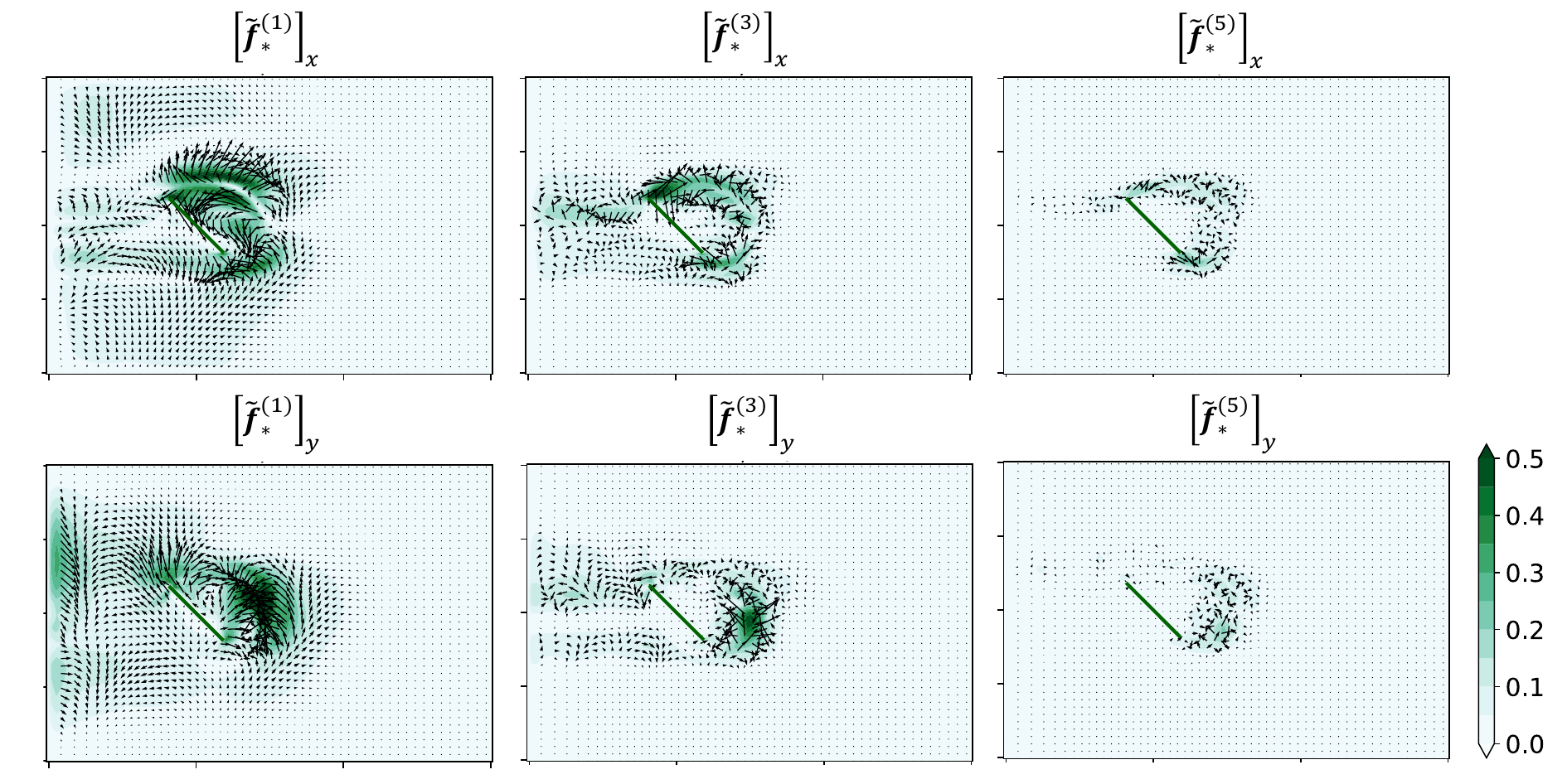}
  \caption{The components of $\tilde{\bm{f}}_{*}^{(m)} (\psi_+=\pi; m=1,3,5)$}
  \label{fig:Fm* (psi+=pi)}
\end{figure}
Figure \ref{fig:Fm* (psi+=pi)} displays the Fourier modes of the components of $\bm{f}_{*} (\psi_+=\pi)$.
We noted that $[\tilde{\bm{f}}_{*}^{(m)}]_y = \tilde{\bm{f}}_{*}^{(m)} \cdot \bm{e}_y$ and so forth.
These vectors represented complex values (c.f. Fig.\ref{fig:Zuv12}).
Owing to the relationships in Eqs. (\ref{eq:symmetry of fe*}), the frequency components were non-zero only for odd values of $m$s.
The major part of $\tilde{\bm{f}}_{*} (\psi_+=\pi)$ was contained in the mode $m=1$,
 and the magnitude decreased as $m$ increased, whereas the spatial homogeneity of the arguments was maintained.
Major upstream part of $[\tilde{\bm{f}}_{*}^{(m)}]_y$ ($\psi_+=\pi$) was contained in the mode $m=1$.

\begin{figure}[h]
  \centering
  \includegraphics[width=0.9\textwidth]{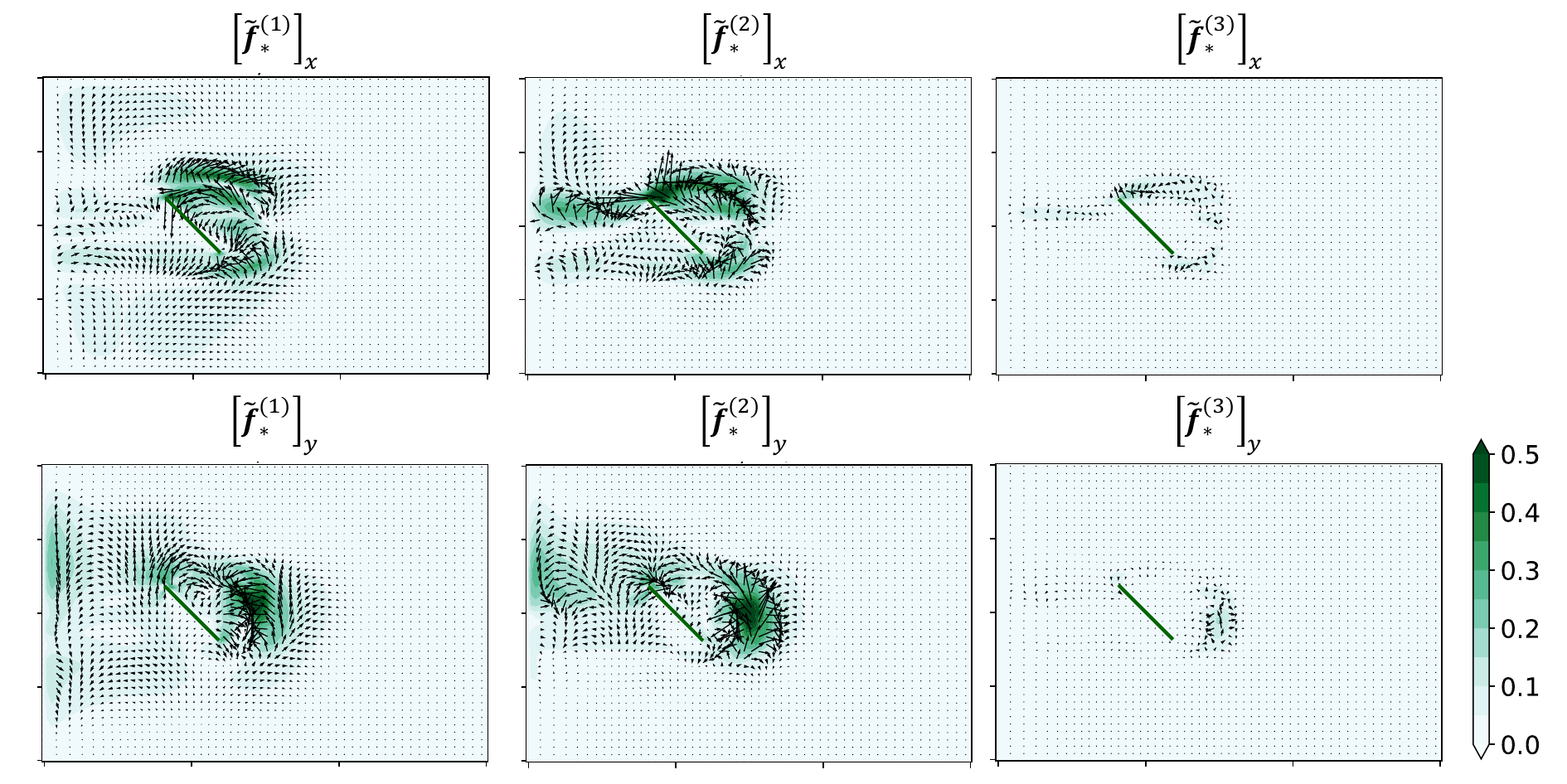}
  \caption{The components of $\tilde{\bm{f}}_{*}^{(m)} (\psi_+=\psi^*; m=1,2,3)$}
  \label{fig:Fm* (psi+=psi*)}
\end{figure}
Figure \ref{fig:Fm* (psi+=psi*)} shows the Fourier modes of the components of $\bm{f}_{*} (\psi_+=\psi^*)$.
The major part of $\tilde{\bm{f}}_{*}^{(m)} (\psi_+=\pi)$ contained in the mode $m=2$ and $1$.
The region upstream of the leading edge exhibited a strong region for $m=2$.

\section{Summary}
In this study, we applied the phase reduction theory to analyze the flow past an inclined plate in a wind tunnel.
The thrust was a result of the lock-in phenomenon cased by the external force.
We employed the Jacobian-free projection method to calculate the phase sensitivity function,
  which allowed for an in-depth analysis of its properties.
Our examination of the frequency decomposition of the phase sensitivity function 
 revealed a prominent component for modes with $k \le 3$.
For mode $k=2$, we observed a strong response region located in the upstream of the leading edge for the plate.

Building on this knowledge, we investigated the lock-in phenomenon induced by a periodic uniform force applied within a rectangular region.
We selected two rectangular regions based on the spatial distribution of the phase sensitivity function
 and determined the optimal position of the rectangle.
Evidently, the optimal position depended on both the mode and the force direction.

Although the primary interactions were observed downstream of the plate,
 the application of the sinusoidal force in the $y-$direction led to an optimal position upstream of the leading edge for the mode $k=2$.
We then compared the lock-in regions with finite amplitude to the predictions of the phase reduction theory.
The numerically calculated Arnold's tongue indicated that for mode $k=1$ in the $y-$direction,
 the result closely matched the theoretical predictions, even for the largest amplitude cases.
This alignment suggests that the theoretical assumption of a linear response remains valid even at higher amplitudes.
Conversely, for the mode $k=2$ in the $y-$direction, the agreement was lost, except for the small amplitude cases.
This discrepancy implies that the linearity does not hold as the amplitude increased.

We applied theories that provide optimal external forces for inducing lock-in phenomena and compared the results with those of a detailed investigation of rectangular external forces.
These optimal forces can result in multiple lock-in phases, which can be beneficial for future control problems.
Furthermore, the optimal forces exhibited distinct features.
Cases with multiple phase lock-ins had multiple strong-force regions, both upstream of the leading edge and downstream of the plate.
Notably, the multiplicity of strong force regions coincides with multiple phase lock-ins.
However, we intend to leave a detailed evaluation of each force regions for future research.

\section*{Appendix}
\subsection{Comparison with the simulation by the spectral element method}
\begin{figure}[h]
  \centering
  \includegraphics[width=0.8\textwidth]{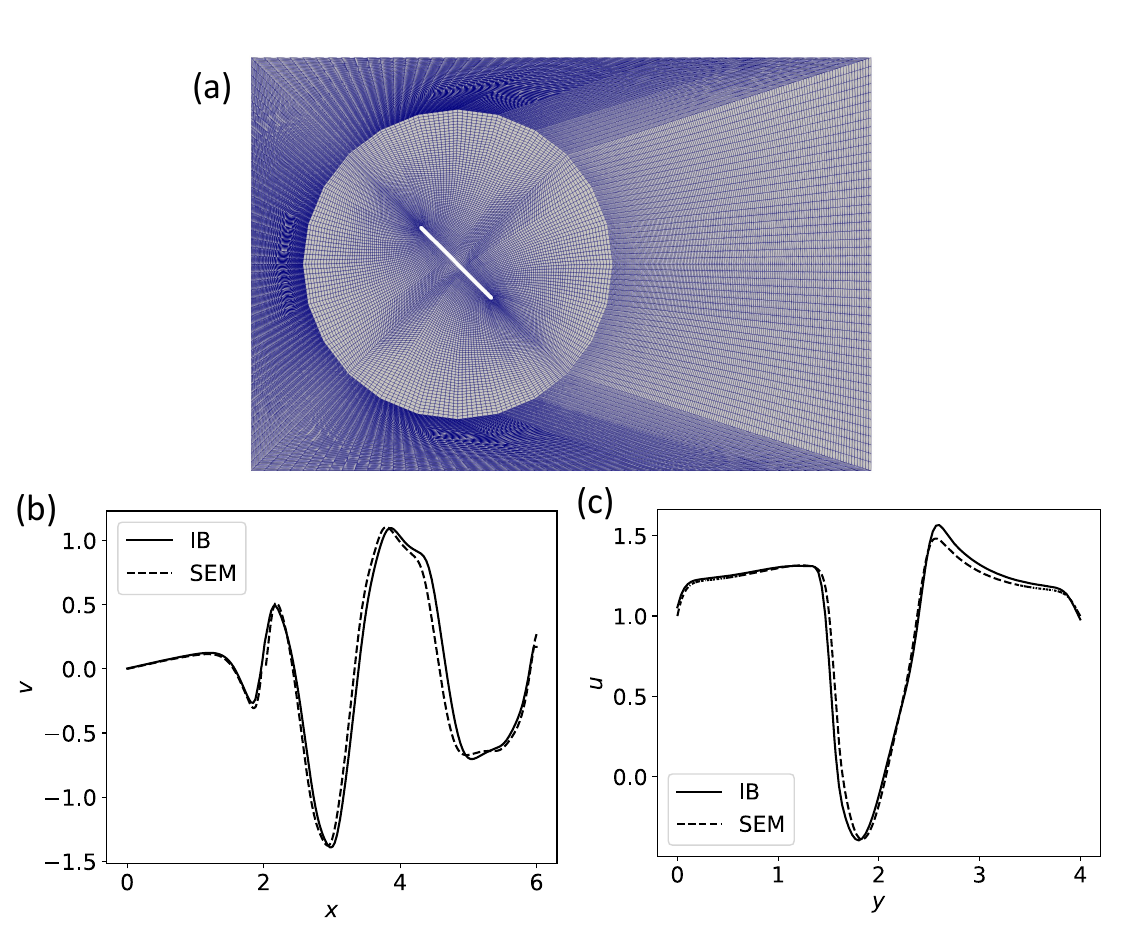}
  \caption{Comparison with the calculation by the spectral element method.}
  \label{fig:Comparison with SEM}
\end{figure}

We compared our calculations with those obtained using the open-source program \textit{Semtex}, which employs the spectral element method (SEM)\cite{blackburn19_semtex, fujita23_inpress}
 to compute the flow around the inclined plate.
In SEM, adaptive elements are used to discretize the plate shape and the computational domain,
 whereas the spectral method is employed to discretize within the elements.
In the calculation, we maintained the same computational domain as that in our main calculation,
 measuring $6 \times 4$.
However, in the SEM, the plate was modeled with a thickness of $t=0.04$, and both edges were represented by half-circles.
The total number of elements in the SEM was 720, and each element was further discretized into $9 \times 9$ elements.
The computational grid is shown in Fig. \ref{fig:Comparison with SEM}(a).
The boundary conditions at the domain boundary remained consistent with those in our calculations, and the time step was set to 0.001.

We obtained a periodic state through a time evolution calculation in SEM, with a period estimated of 3.461. This value exhibited a discrepancy of less than 1 \% compared with the period calculated in our main text by the immersed boundary method (IB), 3.440. 

Additionally, we examined the $v-$fields by IB and SEM along the line connecting two points, $(0,2)$ and $(6,2)$, as shown in Fig. \ref{fig:Comparison with SEM}(b).
Similarly, we observed the $u-$fields along the line connecting two points, $(3,0)$ and $(3,4)$, as shown in Fig.
 \ref{fig:Comparison with SEM}(c).
The results showed a reasonable agreement.
Notably, discrepancies may arise owing to the discretization method and grid spacing.
Furthermore, it is noteworthy that SEM employs a time-evolution calculation, and any differences may be attributed to potential temporal misalignment or a slight error in the selected time.

\subsection{Domain size effect}
\begin{figure}[h]
  \centering
  \includegraphics[width=0.8\textwidth]{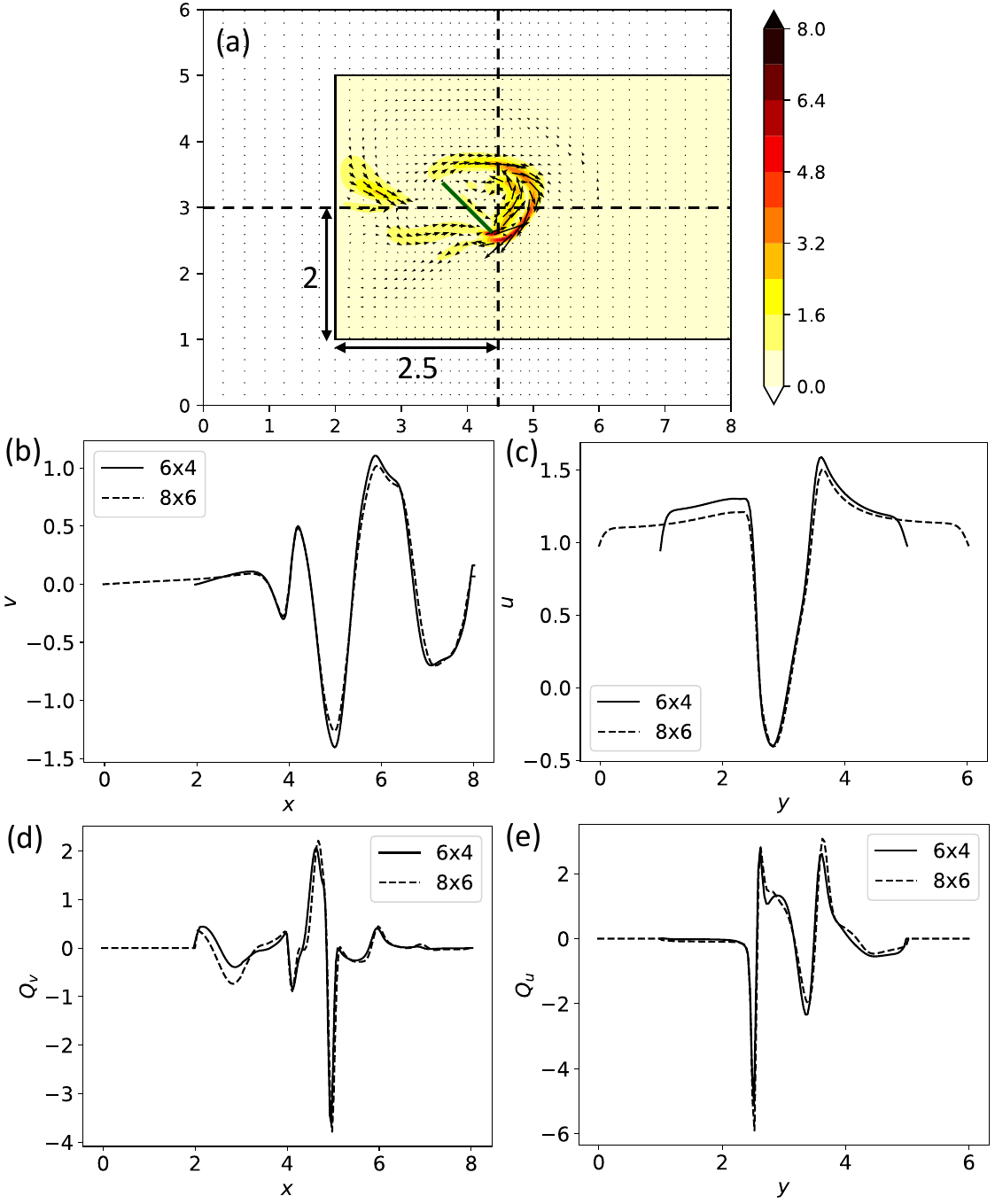}
  \caption{Comparison with different size of the domain.}
  \label{fig:Comparison with different domain}
\end{figure}
We compared the phase sensitivity vectors using different domain sizes.
As shown in Fig. \ref{fig:Comparison with different domain}(a), the $\bm{q}(\bm{x})$ field was computed with a larger computational domain ($8 \times 6$) containing $160 \times 160$ grid points.
To maintain consistency with the main calculation for the computational domain size,
 we employed the projection method\cite{iima21_phase_reduc_techn_target_region} to confine the calculation region to $[2,8]\times[1,5]$.
The center of the plate was positioned at $(4,3)$ to ensure the relative positioning with respect to the domain for the projection method,
 as described in the main text.
The phase in Fig. \ref{fig:Comparison with different domain}(a) is at the origin, matching the upper-left image in Fig.\ref{fig:PSV(aoa=45)}.
Both exhibited similar overall characteristics,
 and the computed Ritz value was -0.0162, which was sufficiently small, despite the computed area being only half of the entire computational domain.

The velocity fields within both domains were compared along two lines, $y=3$ and $x=4.5$, using the coordinate system of the larger computational domain (refer to Fig. \ref{fig:Comparison with different domain}(a)).
The $v$-component along the line $y=3$ and the $u$-component along the line $x=4.5$ are displayed
 in Fig.\ref{fig:Comparison with different domain}(b) and (c), respectively.
The influence of domain size, especially on the width in the $y-$direction, was observed, while the characteristics around the plate remain largely unaffected by the domain.

Figure \ref{fig:Comparison with different domain}(d) and (e) show $q_v$ along the line $y=3$ and $q_u$ along the line $x=4.5$, respectively.
These figures reveal minimal impact of domain near the plate, although slight differences were noticeable in $q_v$ near the upstream (left) boundary.

Overall, these results indicate that the domain size had an insignificant influence on the calculations in this study.

\section*{Acknowledgement}
This work was partially supported by the Japan Society for the Promotion of Science KAKENHI Grant No. 19K03671 and the SECOM Science and Research Foundation.

\end{document}